\documentclass[aps,twocolumn,showpacs,preprintnumbers,nofootinbib,prd,superscriptaddress,groupedaddress,10pt]{revtex4-1}

\makeatletter
\def\l@subsubsection#1#2{}
\def\l@subsubsubsection#1#2{}
\makeatother

\usepackage{graphicx,amssymb,amsmath,amsthm,amsfonts,epsfig,epsf}
\usepackage[usenames]{color}
\usepackage{epstopdf}
\usepackage[caption=false]{subfig}

\usepackage{bm}
\usepackage{dcolumn}
\usepackage{latexsym}
\usepackage{rotating}
\usepackage{longtable}

\usepackage{enumerate}
\usepackage[inline]{enumitem}
\usepackage{tensor,multirow}
\usepackage{url}
\usepackage[linktocpage]{hyperref}

\def\nn{\nonumber}

\newcommand{\tonde}[1]{\left(#1\right)}
\newcommand{\quadre}[1]{\left[#1\right]}
\newcommand{\graffe}[1]{\left\{#1\right\}}
\newcommand{\covar}[2]{{#1}_{#2}}

\newcommand{\dd}{\mathrm{d}}

\newcommand{\modulo}[1]{\left|#1\right|}

\newcommand{\media}[1]{\left\langle#1\right\rangle}

\begin{document}
{\hfill KCL-PH-TH/2018-10}

\title{Impact of multiple modes on the black-hole superradiant instability}
\author{Giuseppe Ficarra}\email{giuseppe.ficarra@kcl.ac.uk}
\affiliation{Dipartimento di Fisica, ``Sapienza'' Universit\`a di Roma \& 
Sezione INFN Roma1, Piazzale Aldo Moro 5, 00185, Roma, Italy}
\affiliation{Department of Physics, King's College London, Strand, London, WC2R 
2LS, United Kingdom}
\author{Paolo Pani}\email{paolo.pani@roma1.infn.it}
\affiliation{Dipartimento di Fisica, ``Sapienza'' Universit\`a di Roma \& Sezione INFN Roma1, Piazzale Aldo Moro 5, 00185, Roma, Italy}
\author{Helvi Witek}\email{helvi.witek@kcl.ac.uk}
\affiliation{Department of Physics, King's College London, Strand, London, WC2R 2LS, United Kingdom}

\begin{abstract}
Ultralight bosonic fields in the mass range $\sim (10^{-20}-10^{-11})\,{\rm eV}$ 
can trigger a superradiant instability that extracts energy and angular 
momentum from an astrophysical black hole with mass 
$M\sim(5,10^{10})M_\odot$, forming a nonspherical, rotating 
condensate around it. So far, most studies of the evolution and end-state of the 
instability have been limited to initial data containing only
the fastest growing superradiant mode. 
By studying the evolution of multimode data in a quasi-adiabatic approximation, we show 
that the dynamics is much richer and depend strongly on the energy of the seed, on the 
relative amplitude between modes, and on the gravitational coupling. 
If the seed energy is a few percent of the black-hole mass, a
black hole surrounded by a mixture of superradiant and nonsuperradiant modes 
with comparable amplitudes might not undergo a superradiant unstable phase, depending on 
the value of the boson mass. 
If the seed energy is smaller, as in the case of an instability triggered by 
quantum fluctuations, the effect of nonsuperradiant modes is negligible. 
We discuss the implications of these findings for current constraints on 
ultralight fields with electromagnetic and gravitational-wave observations.
\end{abstract}

\maketitle

\section{Introduction}

In classical general relativity, where gravity is minimally coupled to massive 
bosonic fields, Kerr black holes~(BHs) can be unstable against the 
superradiant instability (for an overview, see~\cite{Superradiance}). This 
process was discovered almost 50~years 
ago~\cite{zeldo1,Damour:1976kh,Teukolsky:1974yv,Press:1972zz}  but only recently 
it has been subject to intense scrutiny, including rigorous mathematical 
proofs~\cite{Shlapentokh-Rothman:2013ysa,Moschidis:2016wew}.
It was realized that this instability effectively turns 
astrophysical BHs into detectors of axion-like 
particles~\cite{Arvanitaki:2009fg,Arvanitaki:2010sy}
and of ultralight, beyond-standard model bosons in general.

For a BH of mass $M$ and an ultralight boson 
with mass $m_B\equiv \hbar \mu$, the instability is efficient
only when the gravitational coupling $\frac{GM\mu}{c}\sim {\cal O}(1)$, i.e. when the Compton 
wavelength of the particle is comparable to the BH radius. Since astrophysical 
BHs are expected to exist at least in the mass range $\sim (5,10^{10})M_\odot$, 
the superradiant instability is effective for bosons approximately in the mass 
range $m_B\in (10^{-20}-10^{-11})\,{\rm eV}$, i.e. for ultralight bosons. The 
latter are compelling dark-matter candidates\footnote{The collapse and collision 
of compact objects composed of these dark matter candidates has been studied in 
Refs.~\cite{Helfer:2016ljl,Helfer:2018vtq}.} and are predicted in a multitude of 
beyond-standard model 
scenarios~\cite{Jaeckel:2010ni,Essig:2013lka,Hui:2016ltb,Irastorza:2018dyq}. 

The superradiant instability of a Kerr BH has been investigated perturbatively 
for scalar fields~\cite{Detweiler:1980gk,Cardoso:2006wa,Dolan:2007mj,Dolan:2012yt},
including recent exploration of its phenomenological 
implications~\cite{Hannuksela:2018izj,Isi:2018pzk,DAntonio:2018sff,Baumann:2018vus,
Boskovic:2018lkj,Ikeda:2018nhb}, more recently for vector and tensor fields in a 
small-rotation expansion~\cite{Pani:2012vp,Pani:2012bp,Brito:2013yxa,Pani:2013hpa,
Endlich:2016jgc,Cardoso:2017kgn}, and for vector fields around BHs with arbitrary spin 
in an analytical Newtonian 
approximation valid for small gravitational coupling~\cite{Baryakhtar:2017ngi}, 
and numerically for generic values of the BH spin and 
gravitational coupling~\cite{Witek:2012tr,Cardoso:2018tly}, also using
a novel perturbation scheme~\cite{Frolov:2018ezx,Dolan:2018dqv}. Recently, 
nonlinear 
simulations\footnote{The superradiant instability affects also Kerr BHs in 
asymptotically anti de Sitter spacetime; see 
Refs.~\cite{Bosch:2016vcp,Chesler:2018txn,Sanchis-Gual:2015lje} for nonlinear 
simulations in this context.} of the Einstein equations minimally coupled to 
complex single mode
vector fields~\cite{East:2017ovw,East:2017mrj,East:2018glu}
have confirmed the analysis of 
previous quasi-adiabatic and perturbative evolution~\cite{Brito:2014wla} (see 
also Ref.~\cite{Herdeiro:2017phl}).
The latter is justified by the long instability time scale as compared to the
dynamical time scale of the BH.

Note that in the case of complex massive bosonic fields 
or, equivalently, multiple real fields such that the resulting energy-momentum tensor respects 
the symmetry of the Kerr spacetime,
there exist stationary spinning BH solutions surrounded by an oscillating 
condensate~\cite{Herdeiro:2014goa,Herdeiro:2016tmi}. These solutions interpolate 
between boson stars and Kerr BHs and are formed during the evolution of the 
superradiant instability of Kerr BHs against complex 
bosons~\cite{East:2017ovw,Herdeiro:2017phl}. These solutions are unstable 
against higher-order azimuthal modes~\cite{Ganchev:2017uuo} and, at least in 
some region of their parameter space~\cite{Degollado:2018ypf}, the instability time scale is comparable 
to that of Kerr. We deal here with
{\emph{a single real}} bosonic field,
so the only stationary BH configuration is the Kerr metric, as guaranteed by the no-hair 
theorems~\cite{Chrusciel:2012jk,Herdeiro:2015waa}.

The general properties of this process do not depend strongly on the nature of 
the bosonic field: the fundamental unstable mode has a frequency $\omega_R\sim 
\mu$ (henceforth we use $G=c=1$ units) and must satisfy the superradiant 
condition, $\omega_R<m\Omega_H$, where $m$ is the azimuthal number of the 
perturbation and $\Omega_H$ is the BH angular velocity. As a result of the 
instability, a single mode with $m>0$ and arbitrarily small amplitude grows 
exponentially near the BH, extracting energy and angular momentum on a time 
scale $\tau\equiv 1/\omega_I\gg M$, and forming a nonspherical, rotating 
condensate of characteristic size $r_{\rm cloud}\gg M$.
Thus, if ultralight bosonic fields exist in nature, they would produce 
two generic signatures~\cite{Arvanitaki:2009fg,Arvanitaki:2010sy,Superradiance}: 
i) they would favor slowly-spinning BHs 
against highly-spinning ones, since BHs would lose their angular momentum over a 
time scale $\tau$ which can be much shorter than the typical BH accretion rate; 
and ii) they would produce a continuous gravitational-wave~(GW) signal at a 
frequency set by the boson mass. The first signature translates into the 
existence of ``gaps'' in the BH ``Regge plane'', i.e., in its spin--mass 
plane~\cite{Arvanitaki:2010sy,Pani:2012vp,Pani:2012bp,Brito:2013yxa,
Brito:2014wla}, whereas the second signature can be directly searched for in 
LIGO/Virgo (and in the future LISA) data, both as isolated resolvable 
sources~\cite{Arvanitaki:2014wva,Baryakhtar:2017ngi,Brito:2017zvb,
DAntonio:2018sff,Isi:2018pzk,Ghosh:2018gaw} or through the GW stochastic background of a 
population of BH-boson condensates~\cite{Brito:2017wnc}.

Most phenomenological studies so far have focused on the idealized case in which 
the BH is initially surrounded by a \emph{single-mode} superradiant seed 
(see Ref.~\cite{Arvanitaki:2010sy}, 
where the generic setup for the evolution of multiple modes has been laid down, although without discussing the 
phenomenology).
However, more realistic configurations are likely to contain a superposition of 
modes, both superradiant (i.e., satisfying the $\omega_R<m\Omega_H$ condition) 
and nonsuperradiant. This is particularly important if the initial seed is due 
to quantum fluctuations, since in that case modes with different values of $m$ 
are expected to be produced with comparable amplitude.

Full-fledged $3+1$ numerical simulations including the backreaction of massive 
scalar~\cite{Okawa:2014nda} or vector 
fields~\cite{Zilhao:2015tya,Witek2018inprep} onto the spacetime employed such 
{\textit{multimode}} initial data  either through explicit superposition or 
mode mixing due to the construction of metric initial data.
These simulations, furthermore, assumed the presence of an appreciable bosonic cloud,
i.e. a condensate of a few percent of the BH mass.
Those are formed naturally via the superradiant evolution with small seeds~\cite{Brito:2014wla}.
External effects such as a binary companion or 
the inspiral and merger of two such BH-condensate systems will cause mode mixing.
The merger remnant would form in an environment containing a single cloud 
with complex multipolar structure; 
see e.g.~\cite{Baumann:2018vus} for work in this direction.

In those cases, the BH was shifted out of the superradiant regime by absorbing 
a counter-rotating mode with sufficiently large amplitude. This essentially 
switches off the superradiant instability, leaving a rotating BH surrounded by 
a slowly decaying bosonic condensate. These results indicate that the presence 
of multiple modes might crucially change the dynamics of the system.
However, it is unclear whether this conclusion would persist for arbitrarily 
small initial seeds. Ideally, one wishes to follow the nonlinear evolution of a 
small initial seed at least for a few instability $e$-folding times, $\tau\sim 10^6 M$ 
(resp.\ $\tau\sim10^4M$), in the most favorable cases for scalars (resp.\ vectors).
These type of simulations are numerically expensive and, hence, only a small 
number of cases with time scales of~$\sim\mathcal{O}\left(10^{3}\right)M$ were 
analyzed.
Instead, a quasi-adiabatic treatment along the lines of Ref.~\cite{Brito:2014wla}
can provide crucial new insight into the evolution of (multimode) massive bosonic clouds surrounding BHs.

That is precisely the goal of this paper: study the impact of multiple modes on 
the evolution of the superradiant instability. As we shall show, the impact 
of an initial mixture of nonsuperradiant and superradiant modes with comparable 
amplitude depends strongly on the energy of the initial seed and on the value of the 
gravitational coupling $M\mu$. If this energy is 
initially much smaller than the BH mass 
(as expected in the most natural scenarios) the effect of multiple
modes is negligible. On the other hand, if the energy is at least a few percent 
of the BH mass and $M\mu\sim{\cal O}(0.1)$ the presence of nonsuperradiant modes might
affect the evolution and quench the instability completely. 
As we shall discuss, this latter scenario is more 
speculative and may comprise only a small fraction of the BH-scalar condensates expected in the universe.

The rest of this paper is organized as follows.
In Sec.~\ref{sec:Setup} we introduce our setup and different multimode models, and calculate their energy and momentum fluxes.
In Sec.~\ref{sec:QuasiAdiabaticEvol} we present the quasi-adiabatic evolution of our systems. We discuss their implications for 
current electromagnetic and GW-based bounds on the mass of axion-like particles in Sec.~\ref{sec:Implications}.
We conclude in Sec.~\ref{sec:Discussion}.

\section{Setup}~\label{sec:Setup}
We focus on the action describing a real scalar field $\Psi$ with mass $m_B=\mu\hbar$ minimally coupled to gravity,
\begin{equation}
	S=\int d^4x\,\sqrt{-g}\left(\frac{R}{16\pi}-\frac{1}{2}\partial_{\mu}\Psi \partial^{\mu}\Psi-\frac{\mu^2}{2}\Psi^2\right),\label{eq:ScalarAction}
\end{equation}
where $g$ is the determinant of the spacetime metric $g_{\mu\nu}$, and $R$ is the Ricci 
curvature scalar. Minimization of this action yields the Klein-Gordon equation 
$\nabla_\mu\nabla^\mu\Psi=\mu^2\Psi$ and Einstein's equations coupled to the stress energy 
tensor $T_{\mu\nu}=\partial_\mu \Psi\partial_\nu 
\Psi-\frac{1}{2}g_{\mu\nu}\left(\partial_\alpha\Psi 
\partial^\alpha\Psi+\mu^2\Psi^2\right)$.

Our setup will be the same as that of Ref.~\cite{Brito:2014wla}. 
In particular, we study the quasi-adiabatic evolution of the instability, i.e., neglecting the backreaction of the 
scalar field 
and instead employing energy and angular momentum balance argument.
Although the total mass of the condensate can reach a few percent of the black hole mass, the stress-energy 
tensor (e.g. the energy density) remains small and our approximation valid as shown 
in~\cite{Brito:2014wla}.
Furthermore, the energy and angular momentum extraction occurs over the instability time 
scale which is much longer than the BH dynamical time scale; this justifies a 
quasi-adiabatic evolution~\cite{Brito:2014wla,Herdeiro:2017phl}.
%
In this regime the dynamics is governed by the scalar-field equation on a fixed Kerr geometry, the mass and spin of which evolve adiabatically through energy and angular momentum fluxes.

The linearized dynamics of a  Klein-Gordon field on the Kerr background with mass $M$ and spin 
$J = aM = \chi M^{2}$ is described by the Teukolsky equation for a spin-$0$ perturbation, 
whose general solution can be written as
\begin{equation}\label{solKG}
\Psi(t,r,\vartheta,\varphi)=\Re\left[\int d\omega e^{-i\omega t+im\varphi}{_0}S_{{l} m \omega}(\vartheta)\psi_{{l} m \omega}(r)\right]\,,
\end{equation}
where a sum over harmonic indices $({l},\,m)$ is implicit, and
$_{s}Y_{{l} m \omega}(\vartheta,\varphi)={}_{s}S_{{l} m \omega}(\vartheta)e^{im\varphi}$ are the spin-weighted spheroidal harmonics of spin weight~$s$ which, for $s=0$, reduce to the scalar spheroidal
harmonics~\cite{Berti:2005gp}. The radial and angular
functions satisfy the following coupled system of differential equations
\begin{align}
{\cal D}_\vartheta[{_0}S] + & 
        \left[ a^2 (\omega^2-\mu^2)\cos^2\vartheta - \frac{m^2}{\sin^2\vartheta} + \lambda\right]{_0}S=0
\,,\nn\\
{\cal D}_r[\psi]          + &
        \left[\omega^2(r^2+a^2)^2-4aMrm\omega+a^2m^2
\right.\nn\\ & \left.
        - \Delta(\mu^2r^2+a^2\omega^2+\lambda)\right]\psi=0
\,,\nn
\end{align}
where for simplicity we omit the $({l},\,m)$ subscripts,
$r_\pm=M\pm \sqrt{M^2-a^2}$ denotes the coordinate location of the
inner and outer horizons, $\Delta=(r-r_+)(r-r_-)$,
${\cal D}_r=\Delta\partial_r\left(\Delta\partial_r\right)$, and
${\cal D}_\vartheta=(\sin\vartheta)^{-1}\partial_\vartheta\left(\sin\vartheta\partial_\vartheta\right)$.

\subsection{Unstable modes}

Imposing appropriate boundary conditions,
namely purely ingoing waves at the horizon and exponential decay of the scalar field at infinity, a quasi-bound solution to the above
coupled system can be obtained numerically, e.g. using continued fractions~\cite{Cardoso:2005vk,Dolan:2007mj} or a shooting method~\cite{Pani:2013pma}. The eigenspectrum contains an infinite, discrete set of complex quasi-bound modes~\cite{Berti:2009kk}, $\omega=\omega_R+i\omega_I$. 
We will consider only fundamental modes with overtone number $n=0$, i.e. eigenfunctions with zero nodes.
In particular, this system admits unstable ($\omega_I>0$) quasi-bound states satisfying 
the superradiant condition $\omega_R<m \Omega_H$~\cite{Detweiler:1980uk,Dolan:2007mj}, 
with $\Omega_H=a/(2Mr_{+})$ being the angular velocity at the event horizon. For
these solutions the eigenfunctions are exponentially suppressed at spatial infinity:
\begin{equation}
\psi(r)\propto \frac{r^\nu e^{-\sqrt{\mu^2-\omega^2}r}}{r} \quad {\rm as} \quad r\to \infty\,,
\end{equation}
where $\nu=M(2\omega^2-\mu^2)/\sqrt{\mu^2-\omega^2}$.  In the
small-coupling limit, $M\mu\ll 1$, these solutions are well approximated by
a hydrogenic spectrum~\cite{Detweiler:1980uk,Dolan:2007mj} with angular dependence governed by the spherical harmonics $Y_{lm}(\theta,\phi)$, 
angular separation constant $\lambda\simeq {l}({l}+1)$, and frequency
\begin{align}
\label{omega}
\omega \sim & \mu - \frac{\mu}{2}\left(\frac{M\mu}{{l}+1}\right)^2
+i \frac{\gamma_{lm}}{M}\left(m\chi - 2\mu r_+\right)(M\mu)^{4{l}+5}
\,,
\end{align}
where we introduced the dimensionless spin parameter $\chi=a/M = J/M^{2}$
and the coefficient $\gamma_{lm}$ is defined by the following relation:
\begin{align}
\gamma_{lm} = & C_{l}\prod_{j=1}^l{\quadre{j^2\tonde{1-\chi^{2}} +\tonde{m\chi-2\mu r_+}^2}}\,,
\end{align}
with 
$C_{l}=\frac{2^{4l+1}\tonde{2l+1}!}{\tonde{l+1}^{2l+4}}\quadre{\frac{l!}{2l!\tonde{2l+1}!}}^2$, $C_{1}=1/48$ 
for the dominant unstable ${l}=1$ mode. 
From Eq.~\eqref{omega}
it is clear that these modes become unstable ($\omega_I>0$) whenever $\omega_R<m 
\Omega_H$, with an instability time scale roughly given by the $e$-folding time, 
$\tau_{lm}=1/\omega_I$, which strongly depends on the gravitational coupling $M\mu$, 
dimensionless spin $\chi$, and quantum numbers $(l,m)$.

The critical value of the spin that saturates the superradiance condition reads
\begin{align}
\label{eq:ACritical}
\chi > & \chi_{\rm crit} \equiv \frac{4m M\mu}{m^2+4 \mu ^2 M^2}
\,.
\end{align} 
In particular, for positive frequencies and spin, $m>0$ (i.e., a mode corotating with the 
BH) is a necessary but not sufficient condition for the instability.

In the small-$M\mu$ limit, the radial eigenfunctions read~\cite{Brito:2014wla,Detweiler:1980gk,Yoshino:2015nsa}:
\begin{equation}
\psi\tonde{r;\mu,\chi,M}\propto\,g_l{(r)},\label{eq:RadialEigenFunction}
\end{equation}
where $g_l(r)$ can be written in terms of Laguerre polynomials:
\begin{equation}
	g_l{(r)}=\tonde{\frac{2rM\mu^2}{l+1}}^l 
\exp\tonde{-\frac{rM\mu^2}{l+1}}L_0^{2l+1}\tonde{\frac{2rM\mu^2}{l+1}}.\label{
eq:RadialDimensionlessFunction}
\end{equation}
The eigenfunction peaks at
\begin{equation}
	r_{\mathrm{cloud}}\sim\frac{l\tonde{l+1}}{\tonde{M\mu}^2}M,\label{eq:RCloud}
\end{equation}
and thus extends well beyond the horizon, where rotation effects can be neglected.

\subsection{Multiple modes}

\begin{table}[b]
\caption{\label{tab:modes}Initial mode configuration considered in this work.} 
 \begin{tabular}{cccc}
  \hline
  \hline
  Model&  \multicolumn{2}{c}{$(l,m)$}  & $\Psi$  \\
  &  Mode-$1$ & Mode-$2$   &  \\
  \hline
  I    & $(1,1)$	& $(1,-1)$	& Eq.~\eqref{eq:Psi1-1} \\
  II    & $(1,1)$	& $(2,2)$	& Eq.~\eqref{eq:Psi1122} \\
  III   & $(1,1)$	& $(2,1)$	& Eq.~\eqref{eq:Psi1121} \\
  \hline\hline
 \end{tabular}
\end{table}

Using an ansatz of the form~\eqref{solKG}, we consider a generic superposition of monochromatic modes as
\begin{equation}
 \Psi = \sum_{lm} A_{lm}  g_l{(r)}\cos\tonde{m\phi-\omega_R t} P_{lm}(\cos\theta) \,,\label{eq:superposition}
\end{equation}
where $P_{lm}$ are the Legendre polynomials.
For concreteness, we focus on the lowest-lying modes with the shortest absorption or instability timescales
beyond the $l=m=1$ mode,
and consider three different cases ((cf.\ Table~\ref{tab:modes}):
\begin{eqnarray}
\Psi&=&\Psi_{11}+A_{1-1}\,g_1(r)\cos\tonde{\phi+\omega_R t}\sin\theta\label{eq:Psi1-1} \,, \\
\Psi&=&\Psi_{11}+A_{22}\,g_2(r)\cos\tonde{2\phi-\omega_Rt}\sin^2\theta \label{eq:Psi1122} \,, \\
\Psi&=&\Psi_{11}+A_{21}\,g_2(r)\cos\tonde{\phi-\omega_Rt}\cos\theta\sin\theta \,, \label{eq:Psi1121}
\end{eqnarray}
where $\Psi_{11}=A_{11}\,g_1(r)\cos\tonde{\phi-\omega_R t}\sin\theta$.
The first case above (dubbed Model~I) corresponds to the superposition of two modes with $l=1$ and $m=\pm1$, respectively. The second one (dubbed Model~II) corresponds to two modes with $l=m=1,2$, whereas the third case (dubbed Model~III) corresponds to the superposition of two modes with the same $m=1$ but $l=1,2$, respectively. Note that $\omega_R\sim \mu$ for all modes when $M\mu\ll1$. 

It is convenient to express the initial amplitudes $A_{lm}$ in terms of the total mass of the condensate and the relative amplitude between modes. By introducing the scalar-condensate mass computed in the flat spacetime approximation (justified in the $M\mu\ll1$ limit~\cite{Brito:2014wla,Herdeiro:2017phl}),
\begin{equation}
M_S=\int{-T^0_0\,r^2\sin\theta\,\dd r\,\dd\theta\,\dd\phi}\,, \label{eq:ScalarCloudMass}
\end{equation}
we obtain
\begin{eqnarray}
 A_{11}^2&=&\frac{1}{32\pi\tonde{1+\lambda_1^2}}\tonde{\frac{M_S}{M}}\tonde{M\mu}^4 \label{eq:Amplitude1-1} \,, \\
 A_{11}^2&=&\frac{1}{32\pi\tonde{1+81\lambda_2^2}}\tonde{\frac{M_S}{M}}\tonde{M\mu}^4 \label{eq:Amplitude1122} \,, \\
 A_{11}^2&=&\frac{1}{8\pi\tonde{4+81\lambda_3^2}}\tonde{\frac{M_S}{M}}\tonde{M\mu}^4 \label{eq:Amplitude1121}\,. 
\end{eqnarray}
for the three above cases, respectively, where we have introduced the relative amplitudes
\begin{equation}
 \lambda_1=\frac{A_{1-1}}{A_{11}}\,,\quad \lambda_2=\frac{A_{22}}{A_{11}}\,, \quad \lambda_3=\frac{A_{21}}{A_{11}}\,.  \label{eq:Ratio}
\end{equation}
Thus, each initial state is defined by $M_S$ and by one of the $\lambda_i$'s, with $\lambda_i\to0$ being the single $l=m=1$ mode limit.

\subsection{Energy and angular momentum fluxes}

\subsubsection{GW emission from the scalar condensate}
The scalar condensate is a source of GWs. Even though the cloud is nonrelativistic, the 
quadrupole approximation does not apply because the emission is 
incoherent~\cite{Brito:2014wla,Yoshino:2015nsa}.\ Indeed, a dipolar scalar condensate 
would emit quadrupolar GWs at a frequency $\omega=2\omega_R\sim 2\mu$, whose wavelength 
$\sim 1/\omega$ is generally smaller than the size of the source, $r_{\mathrm{cloud}}$. 
Thus, computing the GW emission requires a fully relativistic computation using the Teukolsky 
formalism~\cite{Brito:2014wla,Yoshino:2015nsa}. This is performed in 
Appendix~\ref{app:GWs}, we report here the final result.

For Model~I ($l=1$, $m=\pm1$), we get
\begin{align}	
\dot{E}_{GW}&=\frac{1}{160}\frac{1+\lambda_1^4}{(1+\lambda_1^2)^2}\tonde{\frac{M_S}{M}}
^2\tonde{M\mu}^{14} \,,\label{eq:EdotGW1-1} \\
	\dot{J}_{GW}&=\frac{1}{160\,\omega_R}\frac{1-\lambda_1^2}{1+\lambda_1^2}\tonde{\frac{M_S}{M}}^2\tonde{M\mu}^{14}\,, \label{eq:JdotGW1-1}
\end{align}
for the GW energy and angular-momentum fluxes, respectively.
Clearly, if $\lambda_1\to 0$, both expressions reduce to those of the single-mode case with $l=m=1$~\cite{Brito:2014wla}. In the opposite limit, $\lambda_1\gg1$, the $m=-1$ mode dominates. This corresponds to the same energy flux but, from Eq.~\eqref{eq:JdotGW1-1}, the angular-momentum flux has the opposite sign, as expected for a counter-rotating mode. Thus, the angular momentum variation can be negative when $\lambda_1>1$, i.e. when the initial amplitude of the counter-rotating mode is bigger than that of the corotating one.

For Model~II ($l=m=1,2$), we get
\begin{align}
	\dot{E}_{GW}&=\frac{1}{160\tonde{1+81\lambda_2^2}^2}\tonde{\frac{M_S}{M}}^2\tonde{M\mu}^{14}\,, \label{eq:EdotGW1122} \\
	\dot{J}_{GW}&=\frac{1}{160\,\omega_R\tonde{1+81\lambda_2^2}^2}\tonde{\frac{M_S}{M}}^2\tonde{M\mu}^{14}. \label{eq:JdotGW1122}
\end{align}
In this case the angular momentum variation is always positive because both modes corotate with the BH. When $\lambda_2\to 0$ we retrieve the $l=m=1$ single-mode case, whilst for $\lambda_2\to\infty$ both expressions are suppressed to leading order in $M\mu\ll1$.

Finally, for Model~III ($m=1$, $l=1,2$), we get
\begin{eqnarray}
	\dot{E}_{GW}&=&\frac{1}{10\tonde{4+81\lambda_3^2}^2}\tonde{\frac{M_S}{M}}^2\tonde{M\mu}^{14} \,, \label{eq:EdotGW1121} \\
	\dot{J}_{GW}&=&\frac{1}{10\,\omega_R\tonde{4+81\lambda_3^2}^2}\tonde{\frac{M_S}{M}}^2\tonde{M\mu}^{14}\,. \label{eq:JdotGW1121}
\end{eqnarray}
Again, as $\lambda_3\to0$ we obtain the single mode case, whereas if $\lambda_3\to \infty$ both expressions are suppressed to leading order in $M\mu\ll1$.
We always neglect the GW energy flux at the horizon, which is typically subdominant~\cite{Poisson:1994yf}.

\subsubsection{Superradiant evolution of the scalar condensate}

In the quasi-adiabatic approximation, we assume that the energy and angular-momentum fluxes of the condensate at the BH horizon ($\dot{E}_{S},\dot{J}_{S}$) are entirely converted into the growth of the total scalar-cloud mass and angular momentum~\cite{Brito:2014wla}
\begin{eqnarray}
	\dot{E}_{S}&=&\dot{M}_{S}\,,   \label{eq:ScalarEnergyFlux} \\
	\dot{J}_{S}&=&\dot{L}_{S}\,, \label{eq:ScalarAngularMomentumFlux}
\end{eqnarray}
where ${L}_{S}$ is the $z$ component of the angular momentum of the cloud.
The computation of $\dot M_S$ and $\dot L_S$ is performed in Appendix~\ref{app:SRflux}. 
The procedure can be summarized as follows:
\begin{enumerate*}[label={(\roman*)}]
\item include an adiabatic time dependence $A_{lm}\to A_{lm} e^{t/\tau_{lm}}$ in the expression for the eigenfunction $\Psi$;
      Eq.~\eqref{eq:superposition}. 
Note, that this yields a time dependence of the relative amplitudes 
$\lambda_{i}$; cf. Eq.~\eqref{eq:Ratio} and Eqs.~\eqref{lambda1t}-\eqref{lambda3t} below;
\item compute $M_S(t)$ and ${L}_{S}(t)$ and their corresponding time derivative in the $\omega_I\ll\omega_R$ limit;
\item average the final result over several orbital periods, $T=2\pi/\omega_R$, of the scalar cloud. 
\end{enumerate*}
We report here the final result.

For Model~I ($l=1$, $m=\pm1$), we obtain
\begin{align}
	\media{\dot{E}_S}&\sim 2M_S\,\frac{\omega_{11}+{\lambda_1^2}\omega_{1-1}}{1+{\lambda_1^2}}\label{eq:EdotS1-1omegaIpiccolo} \\
	\media{\dot{J}_S}&\sim 2\frac{M_S}{\mu}\,\frac{\omega_{11}-{\lambda_1^2}\omega_{1-1}}{1+{\lambda_1^2}}\label{eq:JdotS1-1omegaIpiccolo}.
\end{align}
where $\media{...}$ is the time average over several orbital periods and we defined $\omega_{lm}\equiv \omega_I$ for a given $(l,m)$. As discussed in detail below, a crucial point is that $\media{\dot{E}_S}<0$ when $\lambda_1$ is sufficiently large, because $\omega_{1-1}<0$.

For Model~II ($l=m=1,2$), we obtain
\begin{align}
	\media{\dot{E}_S}&\sim 2M_S\,\frac{\omega_{11}+81{\lambda_2^2}\omega_{22}}{1+81{\lambda_2^2}}\label{eq:EdotS1122omegaIpiccolo} \\
	\media{\dot{J}_S}&\sim 2\frac{M_S}{\mu}\,\frac{\omega_{11}+162{\lambda_2^2}\omega_{22}}{1+81{\lambda_2^2}}\label{eq:JdotS1122omegaIpiccolo}.
\end{align}

Finally, for Model~III ($m=1$, $l=1,2$), we obtain 
\begin{align}
		\media{\dot{E}_S}&\sim 2M_S\,\frac{4\omega_{11}+81{\lambda_3^2}\omega_{21}}{4+81{\lambda_3^2}}\label{eq:EdotS1121omegaIpiccolo} \\
		\media{\dot{J}_S}&\sim \frac{1}{\mu} \media{\dot{E}_S} \label{eq:JdotS1121omegaIpiccolo}.
\end{align}
As a consistency check, we note that all above expressions reduce to the expected limits 
when $\lambda_i\to0$ or $\lambda_i\to\infty$. 

In the adiabatic approximation, the time dependence of $\lambda_i$ can be obtained from 
Eq.~\eqref{eq:Ratio} and reads
\begin{eqnarray}
 \lambda_1(t)&=&\lambda_1(0)e^{(\omega_{1-1}-\omega_{11})t} \,, \label{lambda1t}\\
 \lambda_2(t)&=&\lambda_2(0)e^{(\omega_{22}-\omega_{11})t} \,, \label{lambda2t}\\
 \lambda_3(t)&=&\lambda_3(0)e^{(\omega_{21}-\omega_{11})t} \,. \label{lambda3t}
\end{eqnarray}

\section{Quasi-adiabatic evolution}~\label{sec:QuasiAdiabaticEvol}
We are now in the position to study the quasi-adiabatic evolution of the BH-scalar condensate in the presence of multiple modes. Using conservation of the total energy and angular momentum, the evolution of the system is described by~\cite{Brito:2014wla}:
\begin{equation}\label{system}
\left\{\begin{array}{r}
        \dot{M}+\dot{M}_S=-\dot{E}_{{\rm GW}}  \\
	\dot{J}+\dot{L}_{S}=-\dot{J}_{{\rm GW}}   \\
	\dot{M}=-\langle{\dot{E}_S}\rangle       \\
	\dot{J}=-\langle{\dot{J}_S}\rangle  
       \end{array}\right.\,.
\end{equation} 
We integrated this simple set of ODEs using {\textsc{Mathematica}} with its built-in 
NDSolve function.
Using the first relation of Eq.~\eqref{system}, we define the residuals of this procedure as 
${\rm Residuals} \equiv \dot{M} + \dot{M}_S + \dot{E}_{{\rm GW}}$.
We varied the working precision of the NDSolve function, which corresponds to changing the resolution of the numerical scheme,
and verified that the residuals remain $\lesssim \mathcal{O}(10^{-17})$.
At variance with Ref.~\cite{Brito:2014wla}, in Eqs.~\eqref{system} we neglected mass and angular momentum accretion
by ordinary matter (e.g., from an accretion disk),
since the latter play a marginal role in the evolution of the system and are not crucial 
for our purposes.
Indeed, accretion simply introduces an extra time scale in the problem, associated with 
the Salpeter time, $\tau_{\rm accretion}\sim \sigma_T/(4\pi m_p)\sim 4.5\times 10^7\,{\rm 
yr}$, where $\sigma_T$ and $m_p$ are the Thompson cross section and the proton mass, 
respectively. The BH mass and spin grow through accretion approximately over this time 
scale. Thus, a slowly-spinning BH that does not satisfy the superradiant condition can be 
brought into a superradiant phase through accretion~\cite{Brito:2014wla}. Here, for 
simplicity we consider only systems which at $t=0$ satisfy the superradiant condition; 
including accretion is a straightforward extension that should not affect our overall 
conclusion. Within our framework, the evolution depends on the dimensionless parameters 
$M_{0}\mu\equiv M(t=0)\mu$, 
$\chi_{0}\equiv \chi(t=0)$, where $\chi(t)\equiv J(t)/M(t)^2$ is the dimensionless BH spin parameter,
and $\lambda_{i,0}\equiv\lambda_{i}(t=0)$ (with $i=1,2,3$ depending on the model),
as well as on the initial scalar cloud mass $M_{S0}\equiv M_{S}(t=0)$.
The initial angular momentum of the cloud, $L_{S0}\equiv L_S(0)$ is 
determined in terms of $M_{S0}$, $\lambda_{i,0}$ and $\chi_0$, as discussed in 
Appendix~\ref{app:SRflux}.

For concreteness\footnote{The mass simply sets the scale of the problem and we could have 
chosen 
units such that $M=1$. It is straightforward to consider different values of the BH 
mass by rescaling all dimensionful quantities accordingly. For example, the evolution for 
$m_B=10^{-18}\,{\rm eV}$ and $M=10^7M_\odot$ is equivalent to the evolution for 
$m_B=10^{-12}\,{\rm eV}$ and $M_0=10M_\odot$ after rescaling the time coordinate by a 
factor $10^{-6}$.}, we choose to present 
the numerical results in this section for a BH with mass $M_0= 10^7 
M_\odot$ and consider different cases (see also Table~\ref{tab:cases}):
\begin{itemize}
 \item {\bf Case~A:} The scalar field has a mass $m_B=10^{-18}\,{\rm eV}$, corresponding 
to an initial gravitational coupling $M_0\mu\sim 0.075$. The initial BH spin is either 
$\chi_0=0.8$ or $\chi_0=0.95$. In both cases the BH is initially in a 
superradiant state, $\Omega_H>\mu/m$. The initial seed has $M_{S0}=10^{-9} M_0$. This is 
representative for the case $M_{S0}\ll M_{0}$, which includes seeds due to quantum 
fluctuations.
 \item {\bf Case~B:} This is same as Case~A above but for a seed with larger initial mass 
$M_{S0}=0.025 M_{0}$, which should model a seed of astrophysical origin, since the 
energy of the perturbation is a sizeable fraction of the initial BH mass. This case is 
expected when a scalar cloud is present in the environment in which the BH forms, or 
during the coalescence of a binary BH each one endowed with its own scalar cloud. 
 \item {\bf Case~C:} This is same as Case~B above but for a scalar field with slightly 
larger mass, $m_B=4\times10^{-18}\,{\rm eV}$, corresponding 
to an initial gravitational coupling $M_0\mu\sim 0.3$. In this case we set the initial spin to 
$\chi_0=0.95$ in order to satisfy the superradiant condition initially.
\end{itemize}

Case~A was chosen  to agree with the case considered in 
Ref.~\cite{Brito:2014wla}, whereas Case~C is representative for the initial data evolved 
numerically in Ref.~\cite{Okawa:2014nda}. Note that this case is only marginally 
consistent with our small-coupling approximation, $M\mu\ll1$.
In all cases, the BH and boson-field masses 
correspond to a range that will be accessible by future LISA 
observations~\cite{Arvanitaki:2014wva,Baryakhtar:2017ngi,Brito:2017zvb,Brito:2017wnc}. 

\begin{table*}[t]
\caption{\label{tab:cases} Configurations considered in the adiabatic evolution for Models~I-III.}
\begin{tabular}{cccccc}
\hline
\hline
Case    & $m_B/{\rm eV}$        & $M_0/M_\odot$ & $\mu M_0$     & $J_0/M^2$     & $M_{S0}/M_0$  \\
\hline
A       & $10^{-18}$            & $10^7$        & $0.075$       & $0.8,0.95$    & $10^{-9}$     \\
B       & $10^{-18}$            & $10^7$        & $0.075$       & $0.8,0.95$    & $0.025$       \\
C       & $4\times10^{-18}$     & $10^7$        & $0.299$       & $0.95$        & $0.025$       \\
\hline
\hline
\end{tabular}
\end{table*}

\subsection{Model~I: $l=1$ with $m=\pm 1$}
\subsubsection{Case~A: small initial seed}
\begin{figure}[b]
\centering
\subfloat[Scalar cloud]{\includegraphics[width=0.25\textwidth,clip]{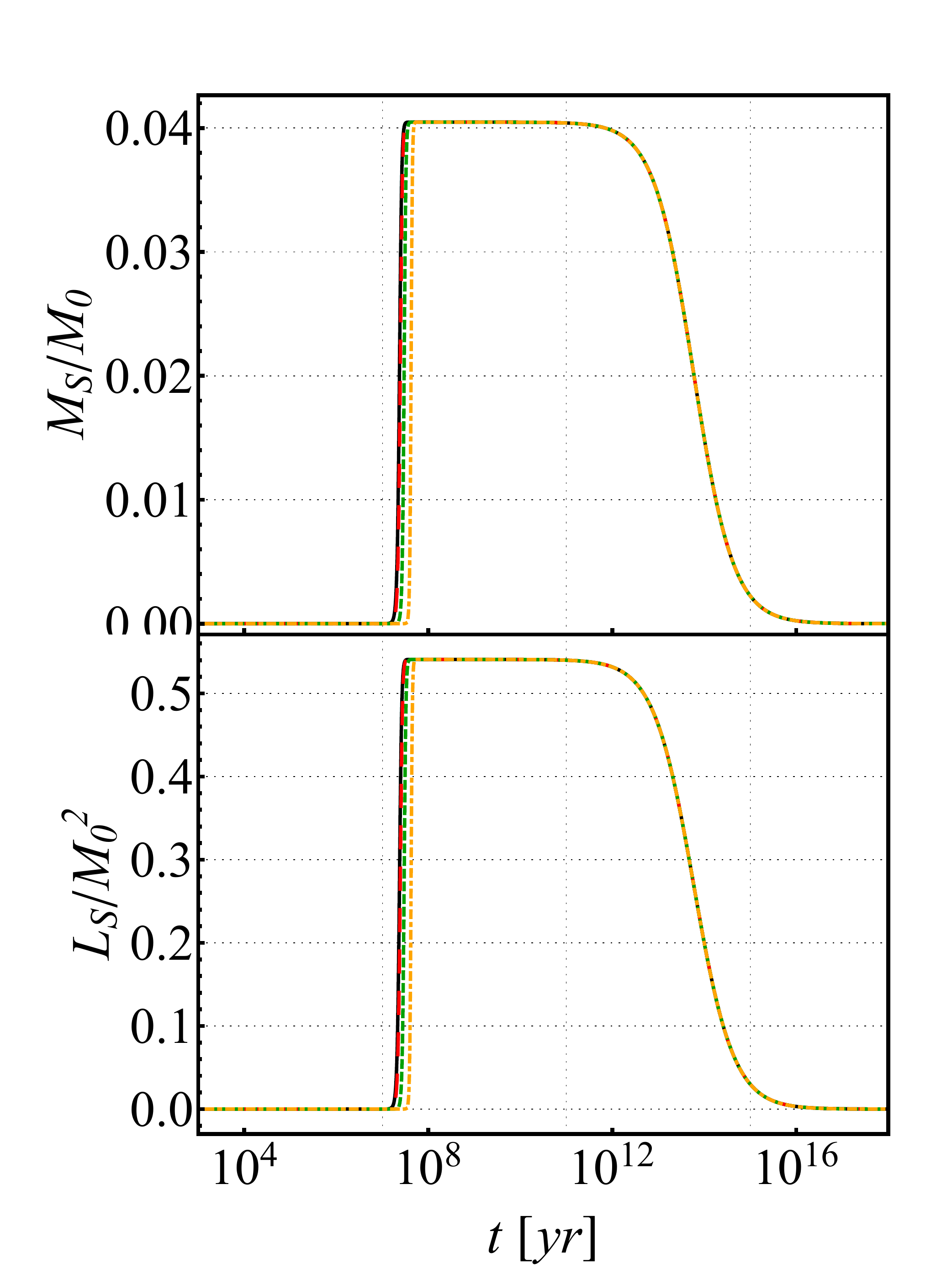}\label{fig:FirstCaseEvolutionSF}}
\subfloat[Black hole]{\includegraphics[width=0.25\textwidth,clip]{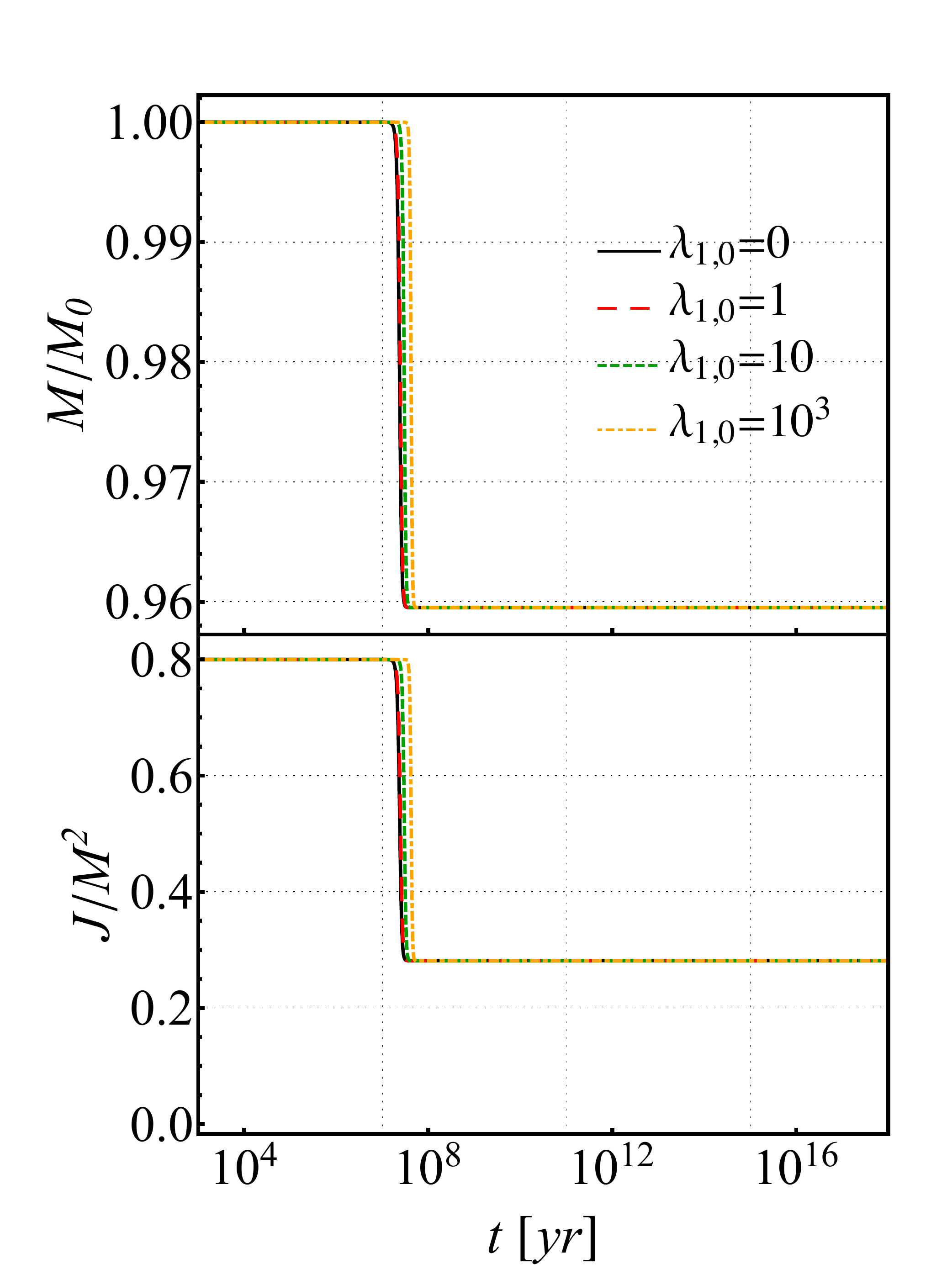}\label{fig:FirstCaseEvolutionBH}}
\caption{\label{fig:FirstCaseEvolution}
Evolution of the~\protect\subref{fig:FirstCaseEvolutionSF} scalar cloud mass (top panel) and angular momentum (bottom panel),
and~\protect\subref{fig:FirstCaseEvolutionBH} BH mass (top panel) and spin (bottom panel) 
for Model~I--Case~A (small initial seeds; cf. Table~\ref{tab:modes}) and different initial relative amplitudes
$\lambda_{1,0}\equiv\lambda_{1}(0)$.
Note that the  evolution is basically insensitive to the presence of a nonsuperradiant mode, i.e. it depend only 
mildly on the relative initial amplitude $\lambda_{1,0}$.
}
\end{figure}
A representative example of the evolution for Model~I in Case~A is presented in 
Fig.~\ref{fig:FirstCaseEvolution}. In this case the evolution is insensitive to the 
presence of a nonsuperradiant unstable mode, even if the latter has initially a much 
larger amplitude than the superradiant mode (e.g., $\lambda_1(0)=10$). This is due to 
Eq.~\eqref{lambda1t}: since $\omega_{1-1}<0$ for a nonsuperradiant mode and 
$\omega_{11}>0$ for the superradiant one, the relative amplitude $\lambda_1(t)$ decreases 
exponentially over a time scale $\sim 1/(\omega_{1-1}-\omega_{11})$.
The evolution is then only affected by the superradiant mode and proceeds as in the 
$l=m=1$ single-mode case~\cite{Brito:2014wla}. For the chosen 
parameters we find $\tau_{11}\sim 7\times 10^6\,{\rm yr}$ which is consistent with the 
exponential growth of the condensate at $t>10^7\,{\rm yr}$ shown in 
Fig.~\ref{fig:FirstCaseEvolution}. In this particular case, the condensate extracts 
$\approx 4\%$ of the initial BH mass. However, its energy density, and hence its backreaction, 
is negligible~\cite{Brito:2014wla} so that it dissipates on a longer time scale through GW 
emission. An estimate for the latter time scale is 
\begin{equation}
 \tau_{\rm GW}\sim \frac{M_S^{\rm max}}{\dot E_{\rm GW}}\sim 6\times10^{13}\,{\rm yr}\,, 
\label{tauGW}
\end{equation}
in agreement with the late-time behavior shown in 
Fig.~\ref{fig:FirstCaseEvolution}.

\subsubsection{Case~B: large initial seed, small coupling}
\begin{figure}[b]
\centering
\subfloat[Scalar cloud]{\includegraphics[width=0.25\textwidth,clip]{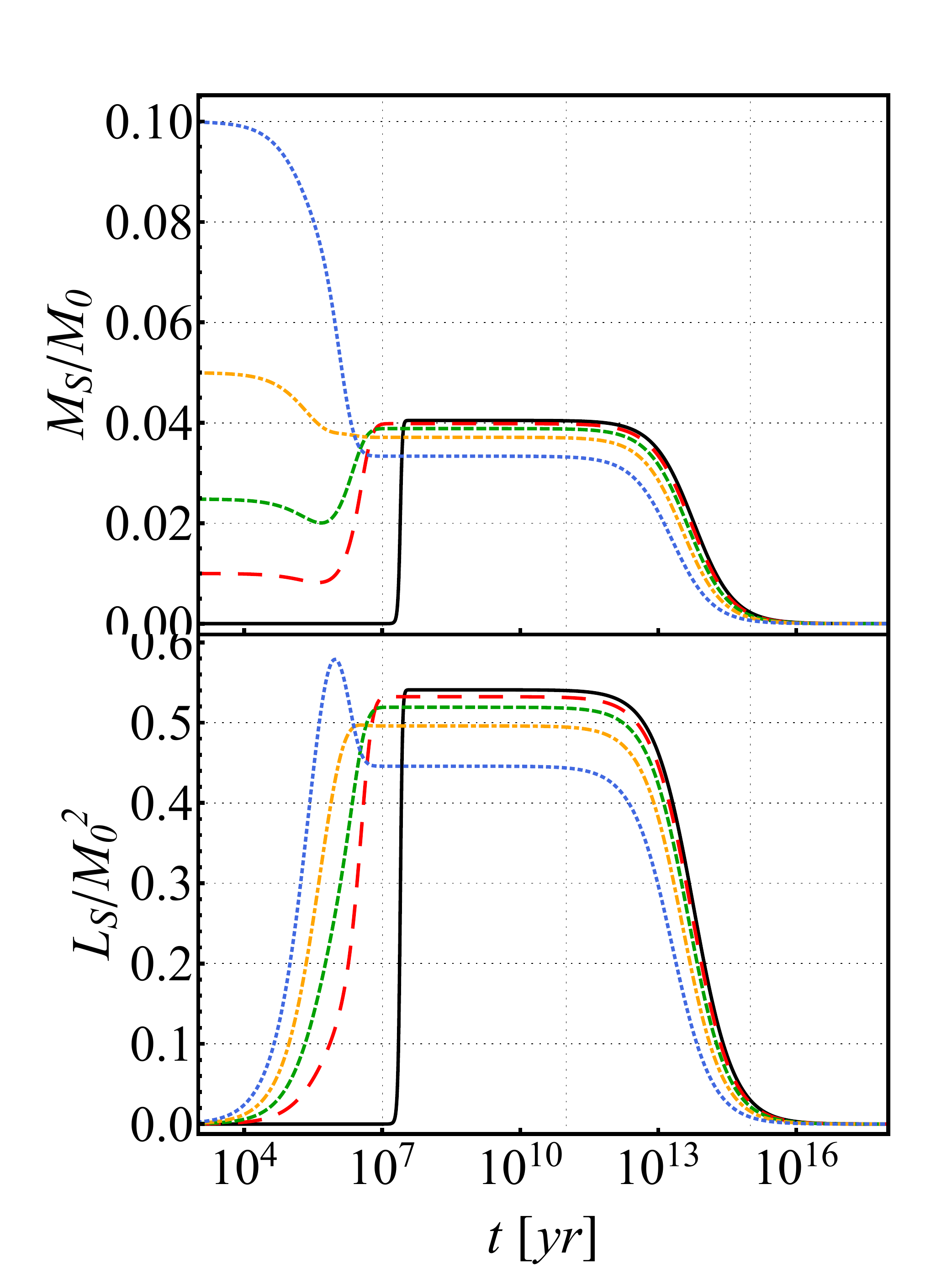}}
\subfloat[Black hole]{\includegraphics[width=0.25\textwidth,clip]{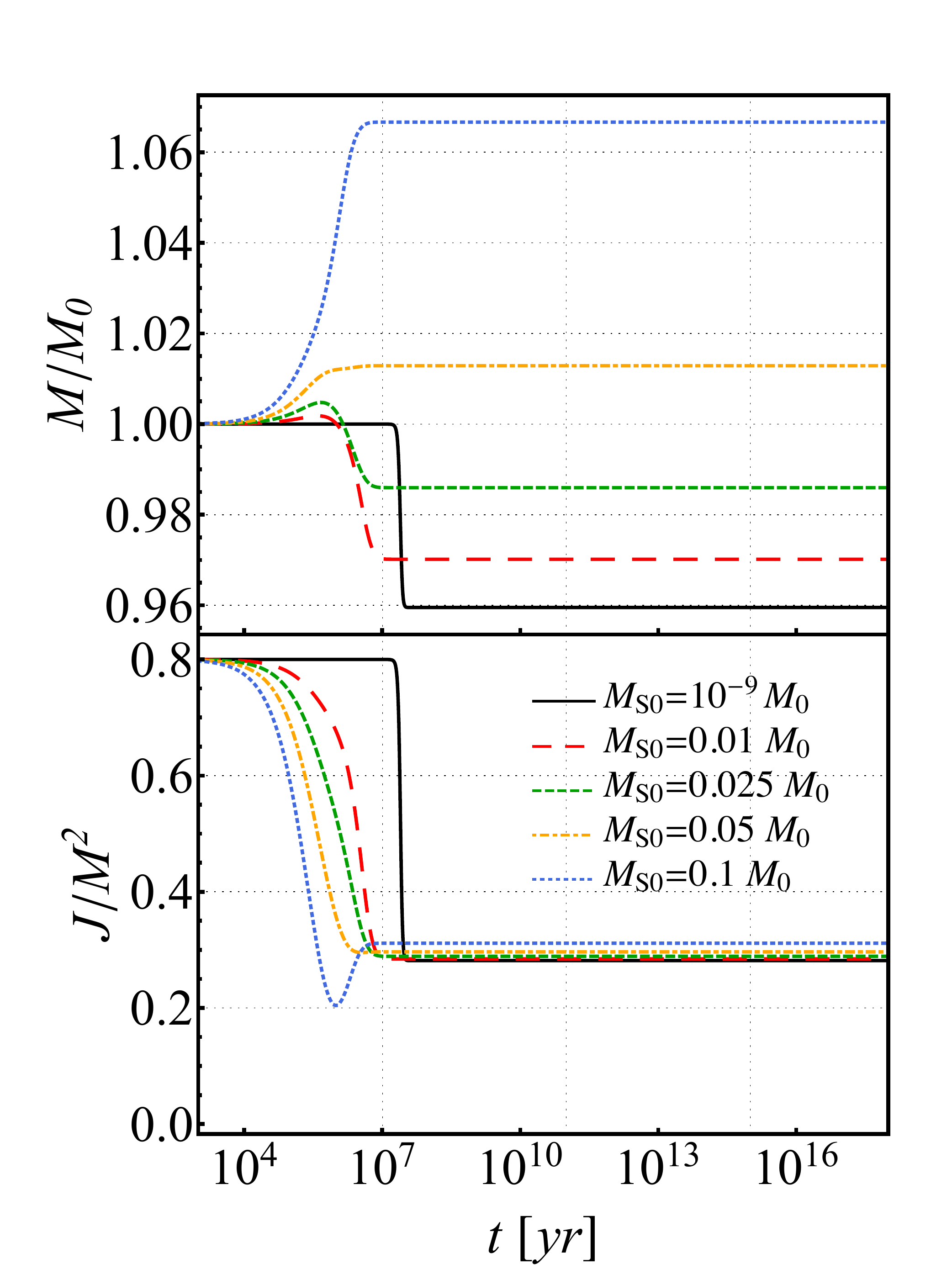}}
\caption{\label{fig:M1largeA}
Same as Fig.~\ref{fig:FirstCaseEvolution} for Model~I, 
but for fixed $\lambda_{1,0}=1$ and different scalar seed amplitudes parametrized by $M_{S0}$.
This includes Case~A (black solid line) and Case~B (green dashed line).
}
\end{figure}

\begin{figure}[b]
\centering
\includegraphics[width=.49\textwidth]{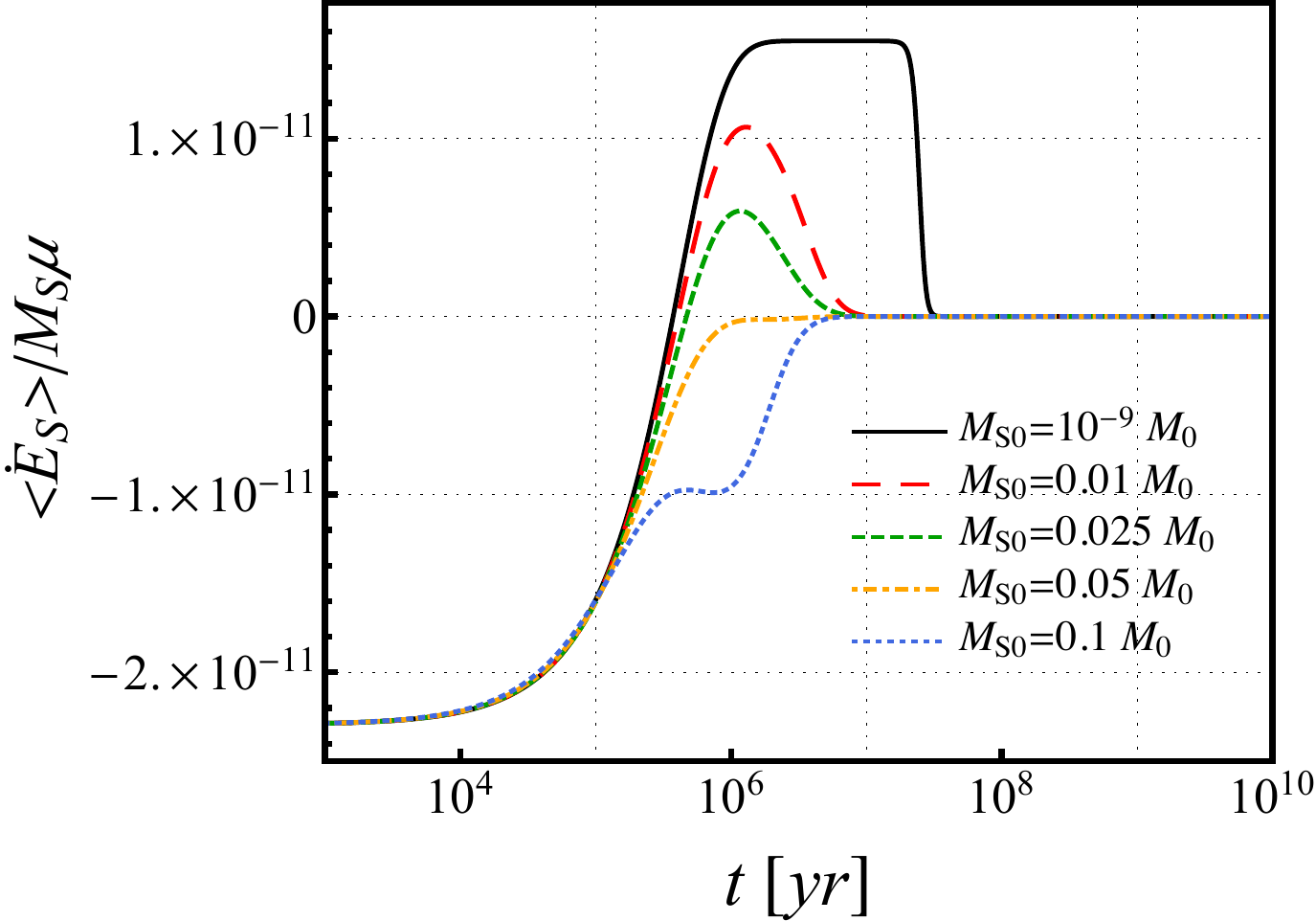}
\caption{\label{fig:Model1flux}
Evolution of the scalar energy flux~\eqref{eq:EdotS1-1omegaIpiccolo}, rescaled by $M_{\rm S}\mu$,
for $\lambda_{1,0}=1$ and different inital scalar cloud masses.}
\end{figure}

The impact of a nonsuperradiant mode is stronger when the initial seed has a larger 
amplitude as illustrated in Fig.~\ref{fig:M1largeA}. There, we show the evolution of 
Model~I
for fixed $\lambda_{1,0}=1$ and different initial scalar cloud masses $M_{S0}$, including
Case~A and Case~B.

The case $M_{S0}=10^{-9} M_0$ corresponds to the case $\lambda_{1}(0)=1$ shown in 
Fig.~\ref{fig:FirstCaseEvolution} whereas, as $M_{S0}$ increases, we observe further 
features. 
Parts of the scalar cloud are initially absorbed by the BH, whose mass, in turn, grows in time.
This can be understood as follows. Neglecting for the moment GW emission, 
system~\eqref{system} reduces to
\begin{equation}\label{systemNOGW}
\left\{\begin{array}{r}
        \dot{M}=-\langle{\dot{E}_S}\rangle \\
	\dot{J}=-\langle{\dot{J}_S}\rangle \\
	\dot{M}_S=\langle{\dot{E}_S}\rangle \\
	\dot{L}_{S}=\langle{\dot{J}_S}\rangle
       \end{array}\right.\,,
\end{equation} 
and therefore, when $\langle{\dot{E}_S}\rangle<0$ and $\langle{\dot{J}_S}\rangle<0$, the 
mass and angular momentum of the condensate decrease 
while the BH mass and spin increase, respectively. From 
Eqs.~\eqref{eq:EdotS1-1omegaIpiccolo} and~\eqref{eq:JdotS1-1omegaIpiccolo}, this can never 
happen when $\lambda_1=0$, because in that case $\langle{\dot{E}_S}\rangle, 
\langle{\dot{J}_S}\rangle \propto \omega_{11}$ and the scalar fluxes are positive 
in the superradiant phase (i.e., when $\omega_{11}>0$). In this case the instability halts 
as the superradiant condition is saturated (i.e., as $\Omega_H\to \mu/m$ or, equivalently, 
as $\omega_{11}\to 0$).
However, the situation is different when $\lambda_1\neq0$. When $\lambda_1\ll1$, 
Eqs.~\eqref{eq:EdotS1-1omegaIpiccolo} and~\eqref{eq:JdotS1-1omegaIpiccolo} reduce to
\begin{align}
	\media{\dot{E}_S}&\sim 2M_S\left[ 
\omega_{11}-(\omega_{11}-\omega_{1-1})\lambda_1^2\right] +{\cal 
O}(\lambda_1^4)\,,\label{eq:EdotS1-1omegaIpiccoloA} \\
	\media{\dot{J}_S}&\sim 2\frac{M_S}{\mu}\left[ 
\omega_{11}-(\omega_{11}+\omega_{1-1})\lambda_1^2\right] +{\cal 
O}(\lambda_1^4)\,,\label{eq:JdotS1-1omegaIpiccoloA}
\end{align}
and therefore the energy and angular-momentum fluxes are smaller than in the single-mode 
case as long as $\omega_{11}>|\omega_{1-1}|$. 

On the other hand, if $\lambda_{1,0}$ is sufficiently large, it might happen that
$\omega_{11} + \lambda^{2}_{1}\omega_{1-1}<0$ and, therefore, the scalar energy flux is negative;
see Eq.~\eqref{eq:EdotS1-1omegaIpiccolo}.
Even when this happens, Eq.~\eqref{lambda1t} shows that $\lambda_1(t)$ decreases 
exponentially. As shown in Fig.~\ref{fig:Model1flux} (solid black curve) when the initial 
seed mass is negligible the scalar flux can be negative at $t=0$, but then it 
turns positive (on a time scale time scale $1/|\omega_{1-1}-\omega_{11}|$) as 
$\lambda_1(t)\to0$.
Therefore, the usual superradiant evolution is not affected by the presence of a 
nonsuperradiant mode as long as the initial seed mass is small.

Larger scalar cloud masses, however, enhance the negative energy flux.
This dependence is illustrated in Fig.~\ref{fig:Model1flux}, where we show the energy flux (rescaled by $M_{\rm S}\mu$)
for $\lambda_{1,0}=1$ and different initial cloud masses. 
In particular, clouds with intermediate values of $M_{S0}$ such as Case~B 
(green dashed line in Figs.~\ref{fig:M1largeA} and~\ref{fig:Model1flux})
may be partly absorbed, but despite a small increase in the BH mass and decrease of the BH spin 
the energy flux becomes positive, i.e., 
the system reaches the superradiant regime.
This picture changes dramatically for larger cloud masses $M_{S0}$;
see, e.g., blue dotted line in Figs.~\ref{fig:M1largeA} and~\ref{fig:Model1flux}. In such
case, the negative scalar flux is significantly enhanced, a sizeable fraction of the 
scalar cloud
-- including its counterrotating modes --
is absorbed by the BH and the superradiant phase can be highly suppressed or entirely 
absent.
The details of the evolution sensitively depend on both the initial relative amplitude $\lambda_{1,0}$ and 
scalar cloud mass $M_{S0}$.

Meanwhile, the angular momentum of the scalar cloud grows because $\langle \dot 
L_S\rangle$ remains positive (see Eq.~\eqref{eq:JdotS1-1omegaIpiccolo}).
As a result, the BH angular momentum decreases irrespective of being in the superradiant phase.
Interestingly, the final BH spin seems to be similar to the single-mode case.
However, a careful investigation of the spin evolution presented in Fig.~\ref{fig:JFinalvslambda} 
shows that this is a coincidence.
For large enough $\lambda_{1,0}$ the final spin depends crucially on the initial parameters.

\subsubsection{Case~C: large initial seed, large coupling}
\begin{figure*}[ht]
\centering
\subfloat[Scalar cloud]{\includegraphics[width=0.5\textwidth,clip]{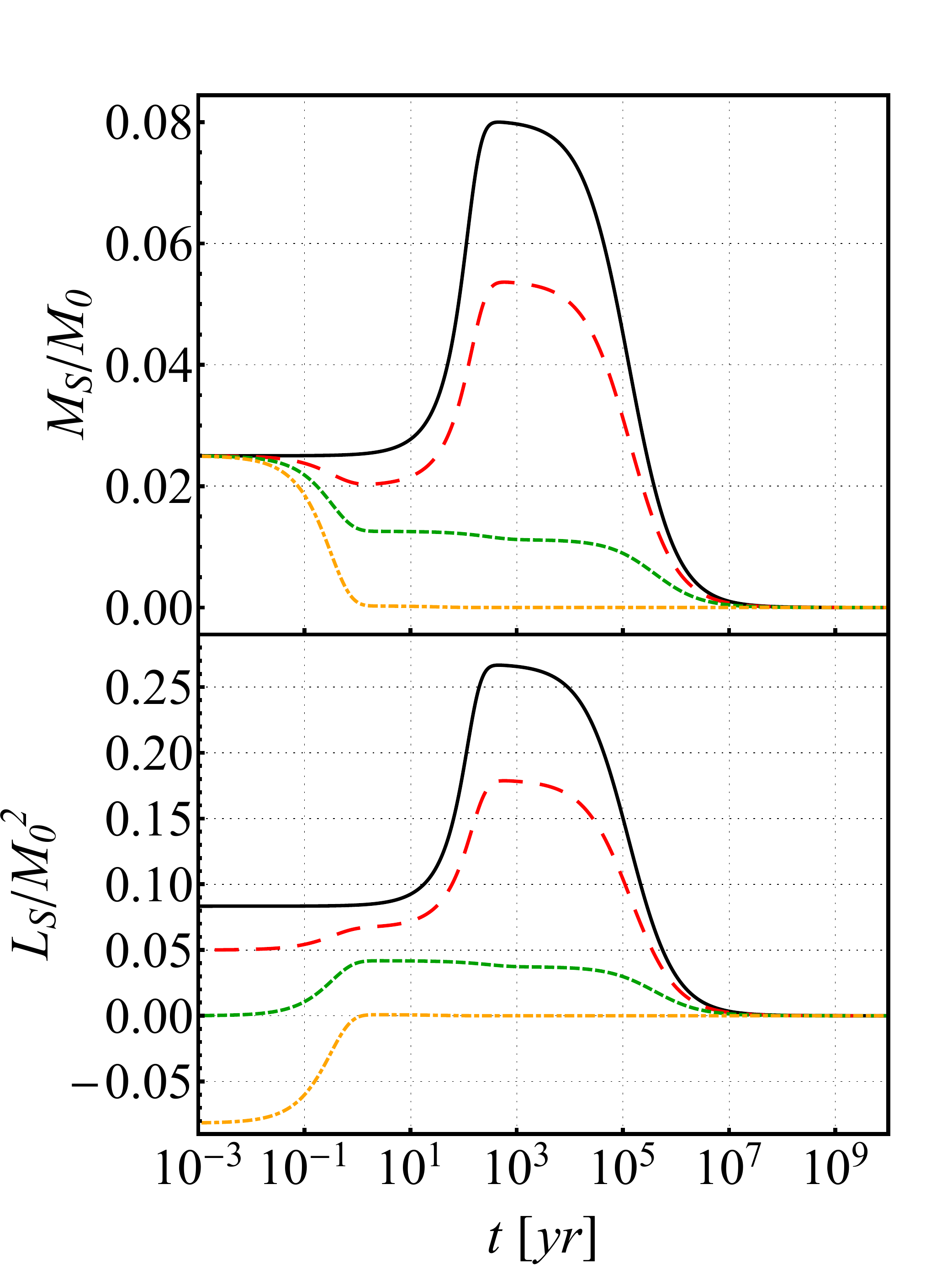}}
\subfloat[Black hole]{\includegraphics[width=0.5\textwidth,clip]{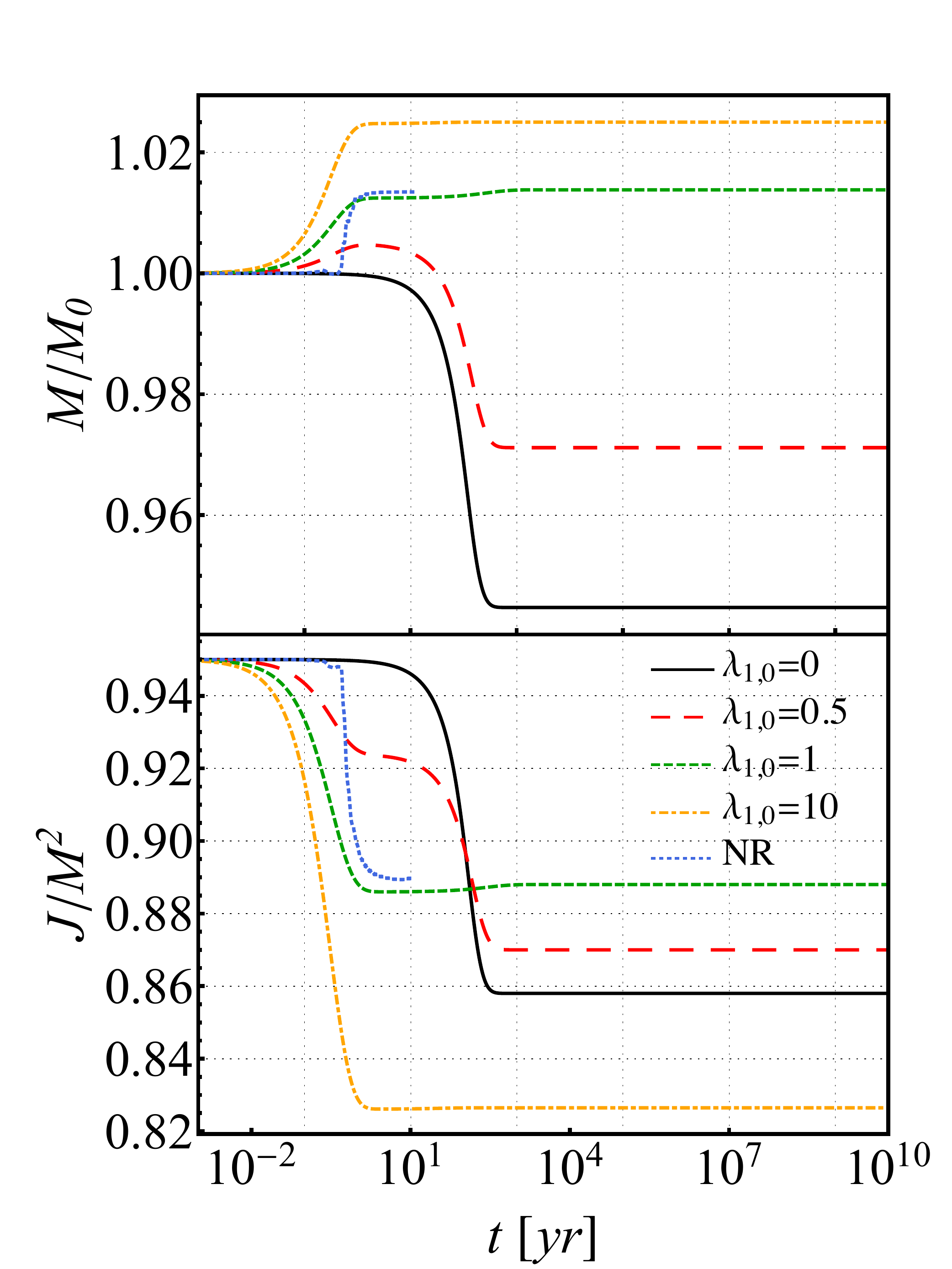}}
\caption{\label{fig:M1largeC} 
Same as Fig.~\ref{fig:FirstCaseEvolution} but for Model~I -- Case~C (cf. 
Table~\ref{tab:cases}),
i.e. $M_{0}\mu=0.3$, $\chi_{0}=0.95$, $M_{S0}=0.025$ and different $\lambda_{1,0}$.
The case $\lambda_{1,0}=1$ (green dased line) has the same initial parameters as the fully nonlinear simulation of~\cite{Okawa:2014nda}
indicated by ``NR'' (blue dotted line). While the BH's early response is different, we recover the same final state.
Note that in the case of large relative amplitude, $\lambda_{1,0}=10$ (yellow dashed lines),
the scalar cloud is completely absorbed on short timescales.
}
\end{figure*}
\begin{figure*}[th]
\centering
\subfloat[$\chi_{0}=0.8,M_{0}\mu=0.075$]{\includegraphics[width=0.50\textwidth,clip]{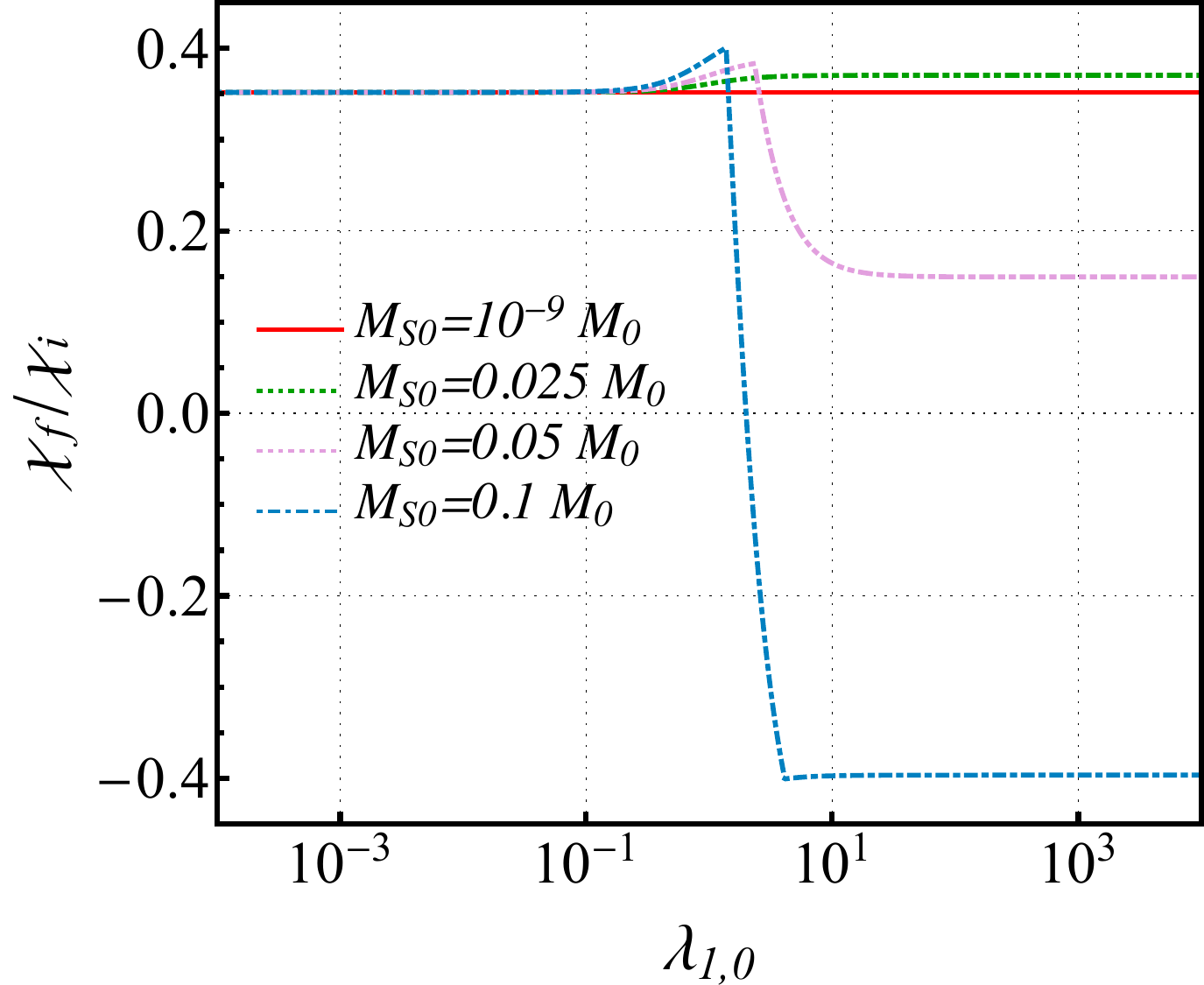}\label{fig:JFinalvslambdaMu0075}}
\subfloat[$\chi_{0}=0.95,M_{0}\mu=0.3$]{\includegraphics[width=0.50\textwidth,clip]{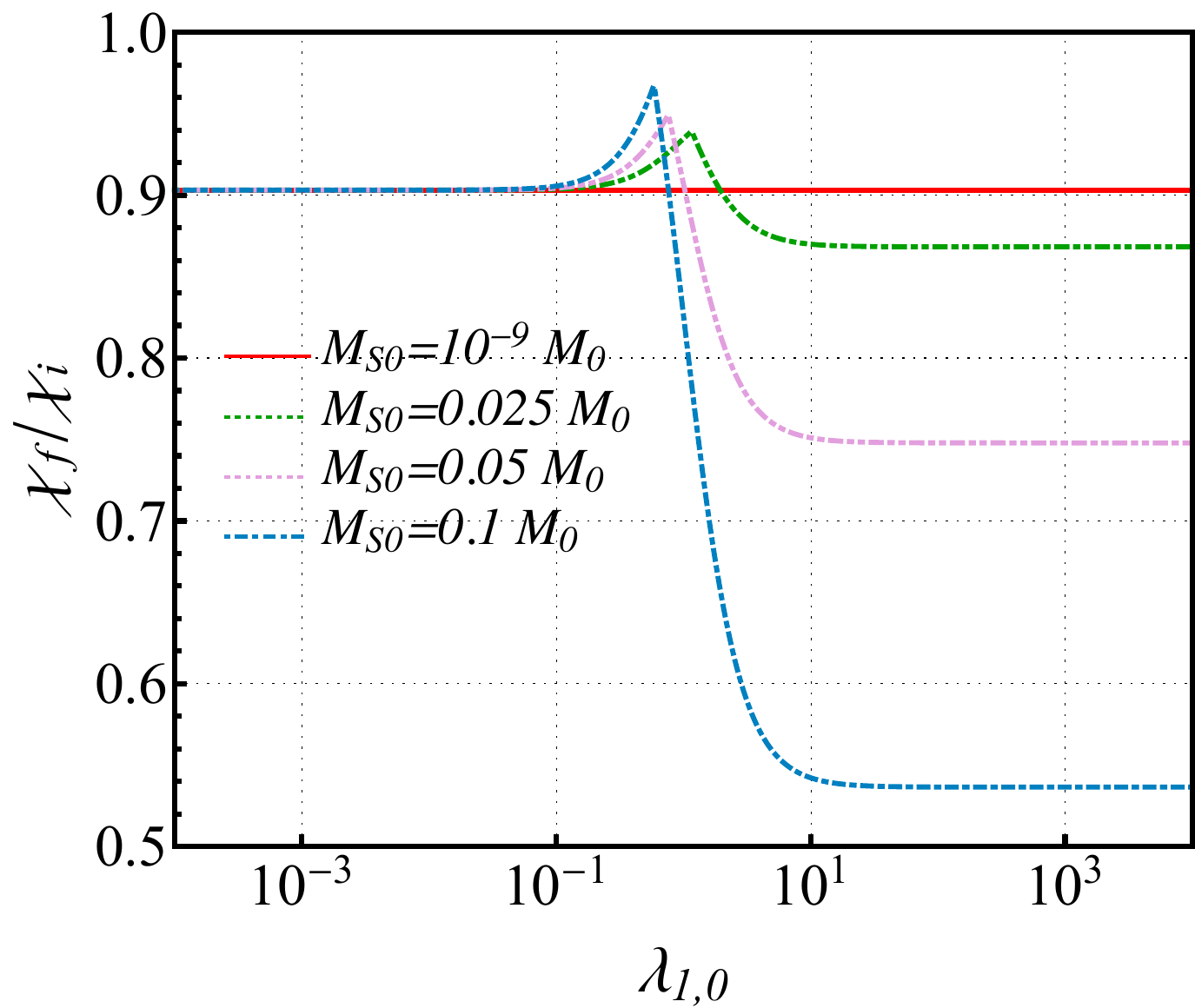}\label{fig:JFinalvslambdaMu03}}
\caption{\label{fig:JFinalvslambda}
Ratio between final and initial dimensionless BH spin as function of the initial relative 
amplitude $\lambda_{1,0}$ of Model~I
and for different initial scalar cloud masses $M_{S0}$.
Fig.~\protect\subref{fig:JFinalvslambdaMu0075} corresponds to Case~A and~B, i.e., has initial parameters ($\chi_{0}=0.8,M_{0}\mu=0.075$).
Fig.~\protect\subref{fig:JFinalvslambdaMu03} corresponds to Case~C, i.e., has ($\chi_{0}=0.95,M_{0}\mu=0.3$).
}
\end{figure*}

When the gravitational coupling $M\mu$ and spin
increase, the impact of the seed mass becomes even more relevant. 
As shown in Fig.~\ref{fig:M1largeC}, the presence of a counterrotating mode with $m=-1$ reduces the superradiant 
energy extraction. For sufficiently large values of $\lambda_{1,0}$ -- in the present case $\lambda_{1,0}\geq1$ -- 
the scalar cloud never grows, since the BH absorbs it before superradiance could kick in.
At the same time, the BH angular momentum decreases because the absorbed energy is mostly contained 
in a \emph{counter-rotating} mode. 
Comparing to the critical value $\chi_{\rm crit}\sim0.882$ 
(see Eq.~\eqref{eq:ACritical}),
we observe that the system can be driven out of the superradiant regime early in the evolution and therefore does not undergo
a superradiant phase.

This behaviour agrees with the expectation raised by fully nonlinear 
simulations~\cite{Okawa:2014nda}.
We show their setup {\textit{KGl\_m30\_a3}} (cf. Table~III of~\cite{Okawa:2014nda})
as blue line in Fig.~\ref{fig:M1largeC}.
While the immediate response differs due to nonlinear (backreaction) effects, we find excellent agreement 
within $\lesssim 0.5\%$ in the BH mass and spin with the adiabatic evolution at late times.

In these cases the final BH spin is not only driven by superradiance but also the absorption of counterrotating
($m=-1$) modes.
Hence, it depends on the initial parameters as illustrated in Fig.~\ref{fig:JFinalvslambda}. 
Here we show the final dimensionless BH spin (relative to its initial value)
as a function of $\lambda_{1,0}$ for different initial cloud masses $M_{S0}$, 
for Cases~A and~B (Fig.~\ref{fig:JFinalvslambdaMu0075})
and Case~C (Fig.~\ref{fig:JFinalvslambdaMu03}).
Small perturbations (red solid lines) always yield the superradiant evolution, i.e., BHs whose final spin is 
smaller than its initial one due to the superradiant instability independently of the presence of a counterrotating mode.
The dependence is more complex for large initial scalar clouds: in an intermediate regime, around $\lambda_{1,0}\lesssim1$
accretion of counterrotating (i.e.,nonsuperradiant) modes and superradiant scattering compete, 
potentially leading to a larger final spin.
Instead, if the initial condensate is dominated by the $m=-1$ mode, the evolution is dominated by accretion
of the counterrotating component and can yield considerably smaller final spins.

{\noindent{\textbf{Summary:}}}
To summarize, our quasi-adiabatic evolution for Model~I reveals that 
the BH superradiant instability proceeds as in the case of a single superradiant mode 
whenever the seed's 
energy is negligible (as in the case of quantum fluctuations), whereas the dynamics and 
the final BH spin are strongly affected by the addition of a nonsuperradiant mode if the 
latter has a large amplitude relative to the superradiant one and if the initial scalar 
cloud has a nonnegligible energy. In some extreme cases, the absorption of the 
counter-rotating superradiant mode is sufficient to reduce the BH angular momentum 
\emph{past} the superradiant condition, so that the instability is completely quenched.

\subsection{Model~II: $l=m=2$ and $l=m=1$}
\begin{figure}[bh]
\centering
\subfloat[Scalar cloud]{\includegraphics[width=0.25\textwidth,clip]{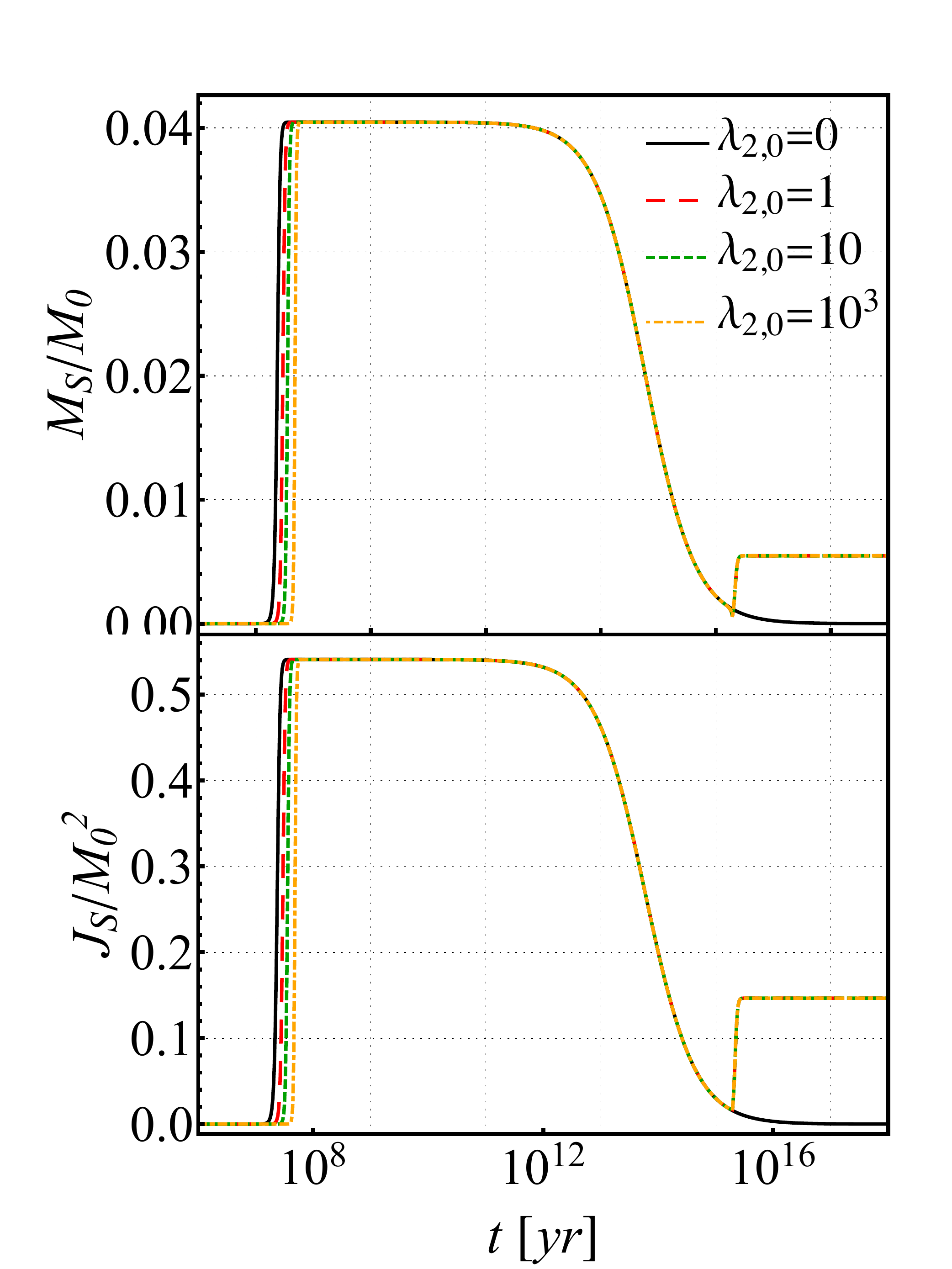}}
\subfloat[Black hole]{\includegraphics[width=0.25\textwidth,clip]{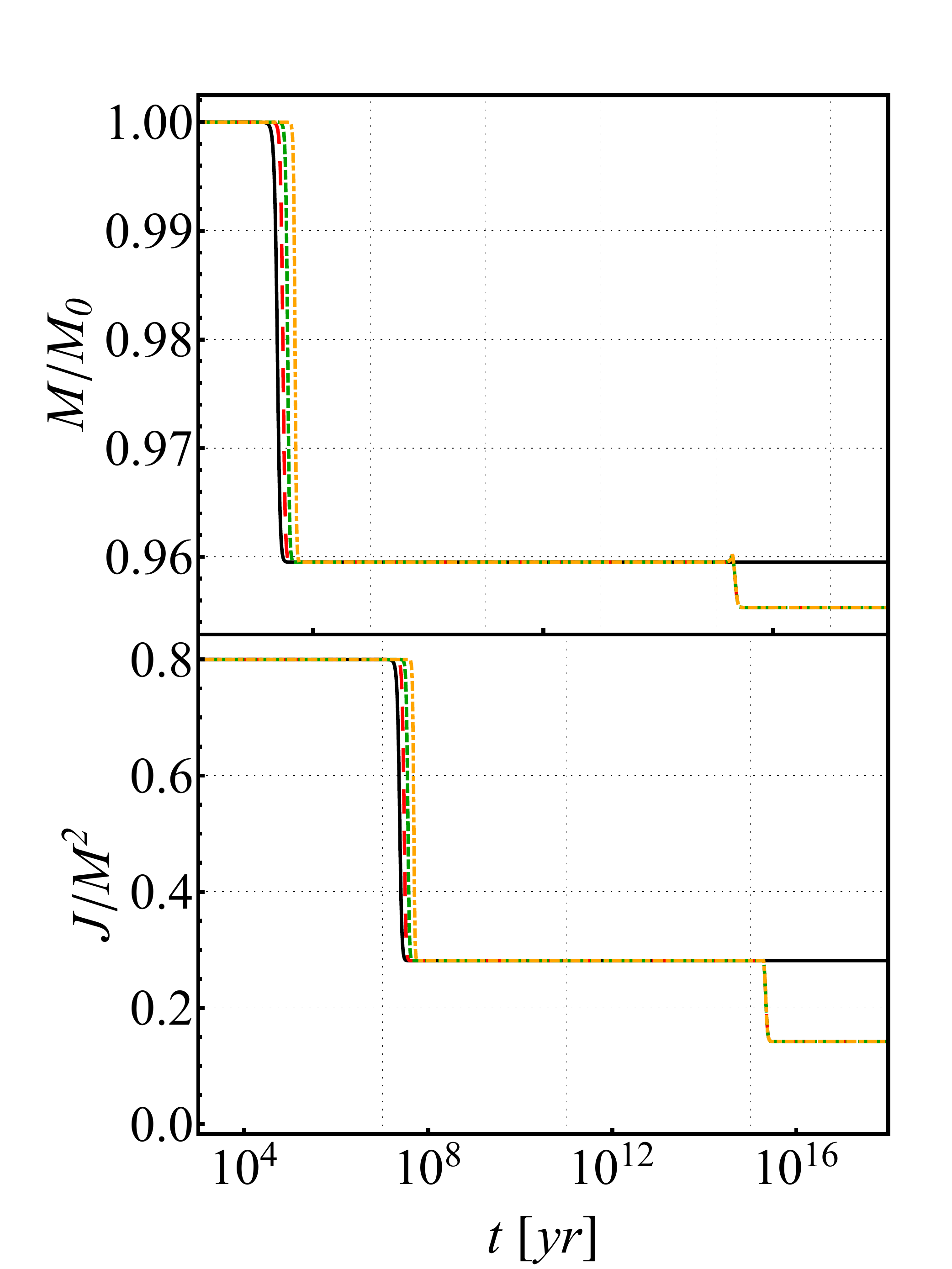}}
\caption{\label{fig:SecondCaseEvolution}
Same as Fig.~\ref{fig:FirstCaseEvolution} but for Model~II (cf. Table~\ref{tab:modes}) and Case~A (small initial seed),
for different values of $\lambda_2(0)$.
We observe a first superradiant growth of the scalar cloud (and extraction of BH mass and angular momentum) 
induced by the dipole mode, followed by dissipation of the cloud due to GW emission.
If the quadrupole mode is present we observe a second superradiant phase independent of the initial relative amplitude
$\lambda_{2,0}\neq0$.
}
\end{figure}
\begin{figure}[th]
\centering
\includegraphics[width=.45\textwidth]{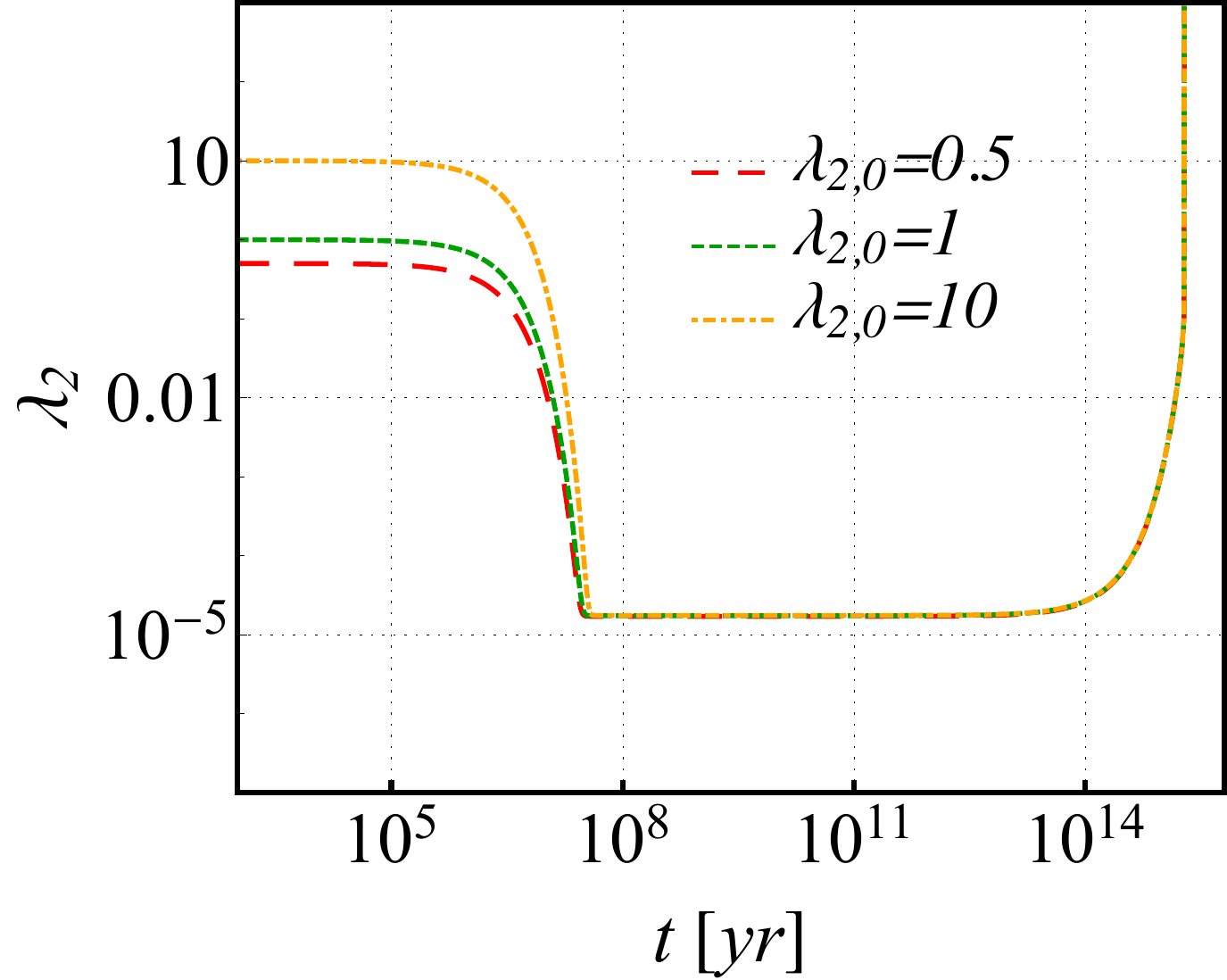}
\caption{\label{fig:lambdavst_ModelII}
Evolution of the relative amplitude $\lambda_2(t)$ for Model~II and 
different initial values $\lambda_{2,0}$. Note that $\lambda_2$ diverges at late times in 
all cases, corresponding to the second superradiant phase of the $l=m=2$ mode.}
\end{figure}

\begin{figure}[bh]
\centering
\subfloat[Scalar cloud]{\includegraphics[width=0.25\textwidth,clip]{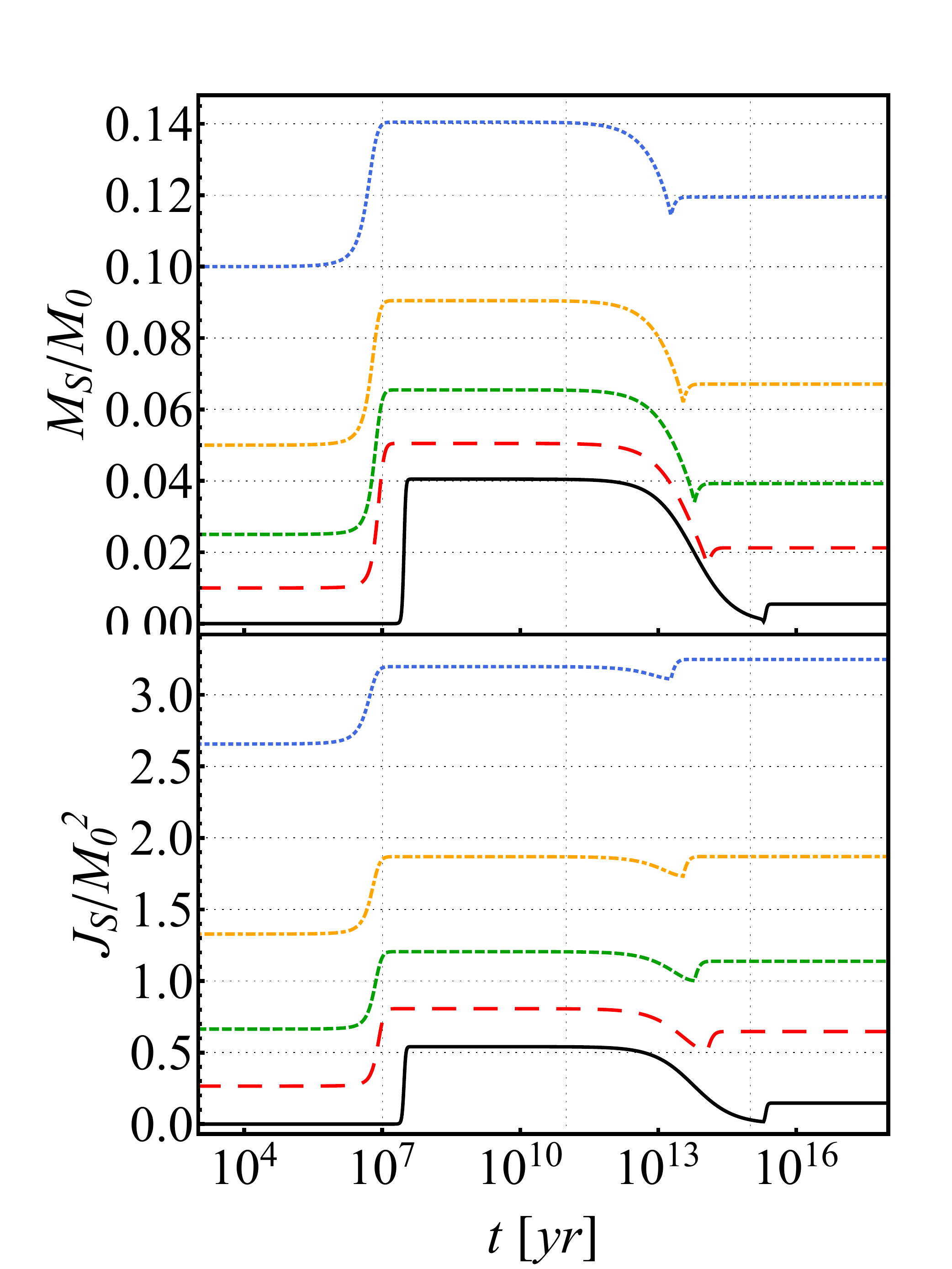}\label{fig:M2largeCScalar}}
\subfloat[Black hole]{\includegraphics[width=0.25\textwidth,clip]{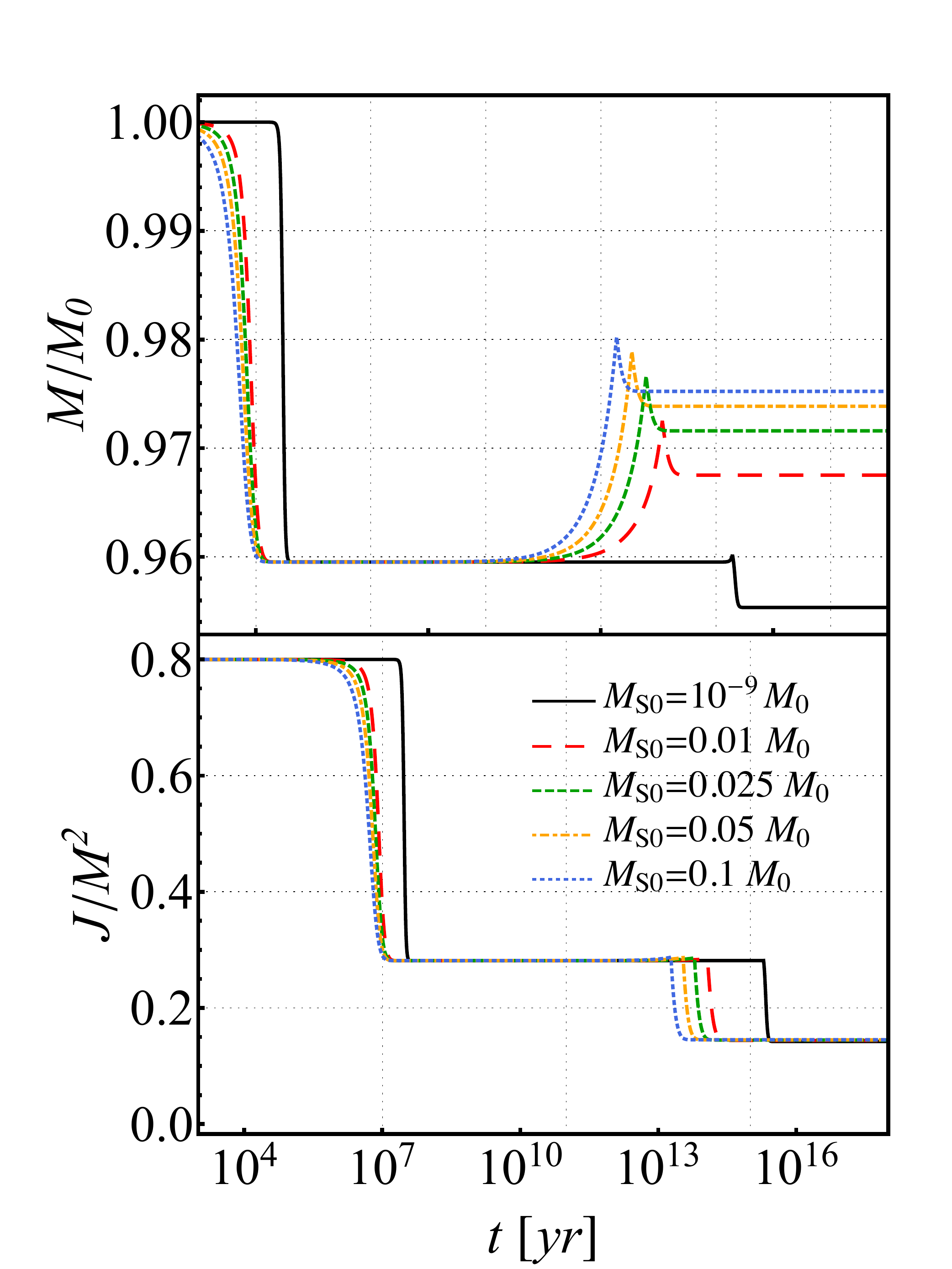}\label{fig:M2largeCBH}}
\caption{\label{fig:M2largeC}
Same as Fig.~\ref{fig:M1largeC} but for Model~II (cf.\ Table~\ref{tab:modes}), i.e.,
we fixed $\lambda_{2,0}=1$ and varied the initial scalar cloud mass $M_{S0}$.
This includes Case~A (black solid line) and Case~B (green dashed line).
}
\end{figure}
\begin{figure}[th]
\centering
\includegraphics[width=.45\textwidth]{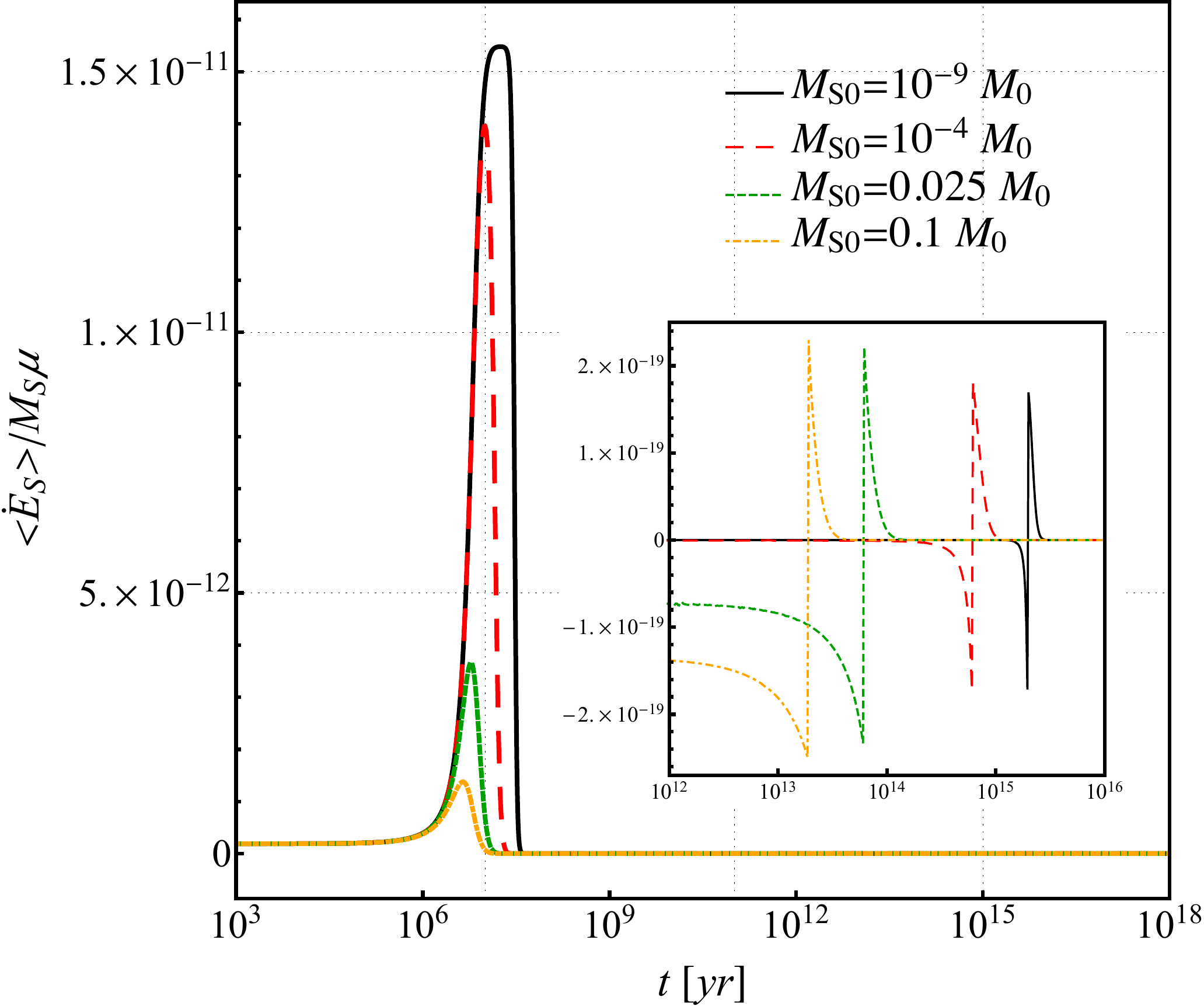}
\caption{\label{fig:Model2flux}
Evolution of the scalar energy flux, rescaled by the scalar cloud mass (and mass parameter),
for Model~II.
The first peak corresponds to the $m=1$ superradiant phase.
During the following dissipation of the cloud the $m=1$ mode decays with $\omega_{11}<0$ still dominating over
the secondary mode. This manifests itself in a {\textit{negative}} energy flux as shown in the inset. Only when the secondary
mode has grown sufficiently the $m=2$ superradiant phase kicks in, as indicated by the second (positive) peak in
the inset. 
}
\end{figure}

The phenomenology of Model~II is different from that of Model~I. 
In particular, both modes can trigger the superradiant instability, albeit on vastly 
different time scales since $\tau\sim (M\mu)^{4l+5}$ depends strongly on $l$.

We first focus on Case~A, i.e. small initial fluctuations, whose evolution is presented in Fig.~\ref{fig:SecondCaseEvolution}
for different relative amplitudes $\lambda_{2,0}$.
We observe two unstable phases: the first one occurring on a time scale 
$1/\omega_{11}$ and the other occurring on longer scales $1/\omega_{22}$.
Because of this separation in time scales,
the evolution starts with the first superradiant phase 
in which the scalar cloud grows and the BH spins down.
Then, the cloud is dissipated through GW emission. 
Finally, the $l=m=2$ mode becomes unstable and the scalar cloud grows 
again, with the BH spin further decreasing since the superradiant threshold $\mu\approx 
m\Omega$ implies a \emph{smaller} final spin; cf.\ Eq.~\eqref{eq:ACritical}.

Looking at Fig.~\ref{fig:SecondCaseEvolution} it is easy to notice that, regardless of 
the value of $\lambda_{2,0}\neq 0$, the end-state of the system remains unchanged, i.e. 
the values of the final BH spin and mass
are an attractor of the dynamics. In order to better 
understand this effect, we study the time evolution of $\lambda_2(t)$, which 
is shown in Fig.~\ref{fig:lambdavst_ModelII}. 
After a first depletion occurring 
at $t\sim \tau_{11}$ due to the superradiant instability induced by the $l=m=1$ mode, 
$\lambda_2$ undergoes an exponential divergence at $t\sim \tau_{22}$. 
This can be seen from the definition of $\lambda_2(t)$, 
Eq.~\eqref{lambda2t}:
At $t\sim\tau_{11}$ the system reaches the superradiant threshold of the $l=m=1$ mode, for 
which $\omega_{11}=0$. 
Afterwards, at $t\sim \tau_{22}$, the secondary mode $l=m=2$ kicks in and takes the system to the superradiant threshold for 
the $m=2$ mode, which is saturated when $\omega_{22}=0$. 
In this situation, however, $\omega_{11}$ becomes negative and $\lambda_{2}(t)$ diverges.

Note that the time scale associated with GW dissipation is much longer for 
$l=m=2$~\cite{Yoshino:2015nsa}, which explains the long time before the condensate 
disappears (not shown in Fig.~\ref{fig:SecondCaseEvolution}).
For the system under consideration, these long time scales force us to consider an 
evolution that lasts much longer than the age of the universe, see 
Fig.~\ref{fig:SecondCaseEvolution}. 
However, if we would consider a stellar-mass BH with $M(0)=10M_\odot$ the time scales would be $10^6$ times 
smaller, since all dimensionful quantities scale with the initial BH mass. 
Then also the secondary superradiant phase might occur within the age of the 
universe.
Model~II--Case~A is therefore a straightforward interpolation between the case of a single mode 
with $l=m=1$ and that of a single mode with $l=m=2$.

We now focus our attention to the influence of a larger initial scalar cloud. 
Its evolution is illustrated in Fig.~\ref{fig:M2largeC} for various $M_{S0}$ and 
equal initial amplitude of the $l=m=1$ and $l=m=2$ modes, i.e., $\lambda_{2,0}=1$.

As before, we observe the growth of the scalar cloud at the expense of the BH mass and angular momentum
on timescales $1/\omega_{11}$, i.e., due to the $l=m=1$ instability.
In contrast to Model~I, this process is essentially independent of the cloud's initial mass since the influence of the secondary mode
kicks in on significantly longer time scales $1/\omega_{22}$.
Once the $m=1$ superradiant threshold is reached, the scalar condensate dissipates via GW emission.
Towards the end of this process, after about $t\sim10^{12}$yr in our setup, the BH mass 
{\textit{increases}};
see Fig.~\ref{fig:M2largeCBH}.
This indicates that the scalar cloud is accreted onto the BH.

To better understand this process, let us inspect the energy flux $\sim\omega_{11} + 81\lambda_{2}^{2} \omega_{22}$;
see Eq.~\eqref{eq:EdotS1122omegaIpiccolo} and Fig.~\ref{fig:Model2flux}. 
Let us also recall the difference in time scales $\omega_{11}\sim (M\mu)^{-9} \gg \omega_{22} \sim (M\mu)^{-13}$.
That is, during the early evolution $\omega_{11}>0$ dominates, thus triggering the $m=1$ superradiant instability
where the scalar flux is positive and peaks around $\tau\sim10^{6}$yr as shown in 
Fig.~\ref{fig:Model2flux}. 
The scalar's growth stops as the superradiant threshold is reached where $\omega_{11}=0$, and the clouds starts dissipating.
Now, although the secondary mode is still growing at a rate $\sim1/\omega_{22}$ (recall 
that $\omega_{22}>0$ is positive), the primary mode
starts decaying with a rate $\omega_{11}<0$. The latter can dominate and lead to a {\textit{negative}}
scalar energy flux as shown in the inset of Fig.~\ref{fig:Model2flux}.
This is consistent with the observation of increasing BH mass in Fig.~\ref{fig:M2largeCBH}.

In the meantime the relative amplitude between the two modes, $\lambda_{2}(t)\sim\exp[(\omega_{22}-\omega_{11})t]$, 
is growing exponentially, see Eq.~\eqref{lambda2t}.
So, eventually the second term in Eq.~\eqref{eq:EdotS1122omegaIpiccolo},
which is positive,
will cancel and then dominate over the $m=1$ contribution.
At this point the scalar energy flux is positive, see inset of Fig.~\ref{fig:Model2flux},
and the BH-scalar cloud configuration undergoes its second (i.e. $m=2$) superradiant phase.
It sets in after about $\tau\sim 10^{13}-10^{15}$yr, with the specific onset depending on 
the 
scalar clous mass; see Fig.~\ref{fig:M2largeCScalar}.
As expected, the scalar cloud grows by dipping into the BH mass and angular momentum that further decreases the 
final BH spin.
Eventually the scalar cloud will dissipate via GW emission on time scales much longer than
shown in Fig.~\ref{fig:M2largeC}.
%

\subsection{Model~III: $l=1,2$ with $m=1$}
Model~III is qualitatively similar to Model~II. 
In particular, this model interpolates between the single-mode case with $l=m=1$ (when 
$\lambda_3\approx0$) and the single-mode case with $l=2$, $m=1$ (when 
$\lambda_3\to\infty$). Also in this case the transition occurs for large values of 
$\lambda_3$, because $|\omega_{21}|\ll|\omega_{11}|$.
The only qualitative difference is related to the critical value of the spin, which is 
the same in both regimes, since the modes have the same azimuthal number $m$ and 
Eq.~\eqref{eq:ACritical} does not depend on $l$.

\section{Implications}~\label{sec:Implications}
Due to BH no-hair theorems for real bosonic fields~\cite{Chrusciel:2012jk,Herdeiro:2015waa}, the end-state of the 
evolution must be a Kerr BH, the condensate being eventually dissipated in GWs. However, an interesting question 
concerns the final value of the BH spin in the new stationary configuration and, more generically, the phase-space 
(Regge-plane) of the final BH.

Another relevant question concerns the viability of our Cases~B~and~C, where the energy of the initial seed is 
a sizeable fraction (roughly $2\%$) of the BH mass. This scenario could occur if the BH is formed in a scalar-rich 
environment, for example if it is formed out of the merger of two previously scalarized BHs. The time scale for GW 
dissipation of the condensate strongly depends on the coupling~\cite{Arvanitaki:2014wva,Brito:2017zvb,Brito:2017wnc}. 
For the fundamental $l=m=1$ mode, 
\begin{equation}
 \tau_{\rm GW}\sim  10^{10}\, \left(\frac{0.5}{\chi}\right)\left(\frac{M}{10^6\,M_\odot}\right)  
\left(\frac{0.1}{M\mu}\right)^{15} \,{\rm yr}\,. \label{tauGWtotal}
\end{equation}
Thus, depending on the mass and spin of the BH and on the mass of the bosonic field, $\tau_{\rm GW}$ can 
easily exceed the age of the universe, in agreement with Eq.~\eqref{tauGW} above. In that case the condensate 
will not have enough time to dissipate during the coalescence, and the merger remnant will form in an environment where 
the energy of the scalar field is not negligible.

Although this scenario might be relevant only for a fraction of sources, the majority of supermassive BHs are 
believed to form via hierarchical mergers. Thus, they constitute a sizeable fraction of the GW signal from bosonic 
condensates which may potentially be detected by LISA~\cite{Brito:2017zvb,Brito:2017wnc}.
We leave a more quantitative analysis of possible formation scenarios and event rates for future work.

%
\subsection{Spin evolution}
{\noindent{\textbf{Single mode:}}}
For reference, let us recall the final spin resulting from the evolution of the single, 
$l=m=1$ mode. 
Focusing on the initial parameters used in our previous sections, the final spin is
$\chi(t\to\infty)\approx 0.28$ and $\chi(t\to\infty)\approx 0.86$ for Case~A and Case~C, 
respectively.

{\noindent{\textbf{Model~I:}}}
The presence of a counter-rotating mode can significantly change the value of the final spin,
and details depend on the initial parameters.
To better understand those dependencies we present 
the ratio between the final and initial spin as function of the initial relative amplitude in Fig.~\ref{fig:JFinalvslambda}.
They are shown for different (initial) scalar cloud masses and fixing
($\chi_{0}=0.8$, $M_{0}\mu=0.075$), i.e., Model~I--Case~A and~B (see 
Fig.~\ref{fig:JFinalvslambdaMu0075})
or
($\chi_{0}=0.95$, $M_{0}\mu=0.3$), i.e., Model~I--Case~C (see Fig.~\ref{fig:JFinalvslambdaMu03}).

If we consider only small fluctuations 
this ratio remains constant, i.e., is insensitive to the presence of a counterrotating mode and yields the
same final spins as in the single ($l=m=1$) mode case reviewed above.

Similarly, the final BH spin appears independent of the initial scalar cloud mass $M_{S0}$ as long as the relative amplitude
between counter- and co-rotating modes
is sufficiently small, namely $\lambda_{1,0}\lesssim0.1$.

However, if the initial scalar cloud contains comparable excitations of the $m=\pm1$ modes, i.e. if $\lambda_{1,0}\sim\mathcal{O}(1)$,
and its mass is a few $\%$ of the BH's initial mass, the dependency of the final spin is more complex.
Now, the angular momentum flux~\eqref{eq:JdotS1-1omegaIpiccolo} is determined by both the superradiant $m=+1$ mode and the
counter-rotating $m=-1$ mode. Since $\omega_{11}>0$ and $\omega_{1-1}<0$, both modes will {\textit{increase}}
the flux and, hence, reduce the final BH spin, although the latter may be slightly larger 
than in the single $l=m=1$ mode case, depending on the parameters (see 
Fig.~\ref{fig:JFinalvslambda}).

As we further increase $\lambda_{1,0}\gtrsim10$ the ratio between the final and initial BH spin approaches
a constant value (i.e., independent of $\lambda_{1,0}$)
that can, in fact, be {\textit{smaller}} than the $l=m=1$ case
but depends strongly on the the initial scalar cloud mass $M_{S0}$.
Note, that the spin can actually flip sign due to absorption of counter-rotating modes as shown in Fig.~\ref{fig:JFinalvslambdaMu0075}.

{\noindent{\textbf{Model~II:}}}
Due to the presence of a secondary superradiant mode, the extraction of the BH spin 
proceeds in two stages:
the first one is caused by the dipole mode and yields the same final BH spin as the single 
($l=m=1$) mode case 
discussed above. After $t\sim1/\omega_{22}$ the system undergoes the $l=m=2$ superradiant phase that yield further extraction of the 
BH spin. As indicated in Figs.~\ref{fig:SecondCaseEvolution} and~\ref{fig:M2largeC},
its final value appears independent of the initial relative amplitude $\lambda_{2,0}$ or scalar cloud mass $M_{S0}$.
This is because $\lambda_{2}(t)$ will have acquired the same value (independent of its 
initial one) by the time the $l=m=2$ mode becomes active; see 
Fig.~\ref{fig:lambdavst_ModelII}.
For example, in the case studied here (with $\chi_{0}=0.8$), the final spin is $\chi(t\to\infty)\sim 0.14$.

\subsection{Regge planes}
\begin{figure*}[th]
\centering
\subfloat[$M_{S0}=10^{-9}M_{0}$]{\includegraphics[width=0.33\textwidth,clip]{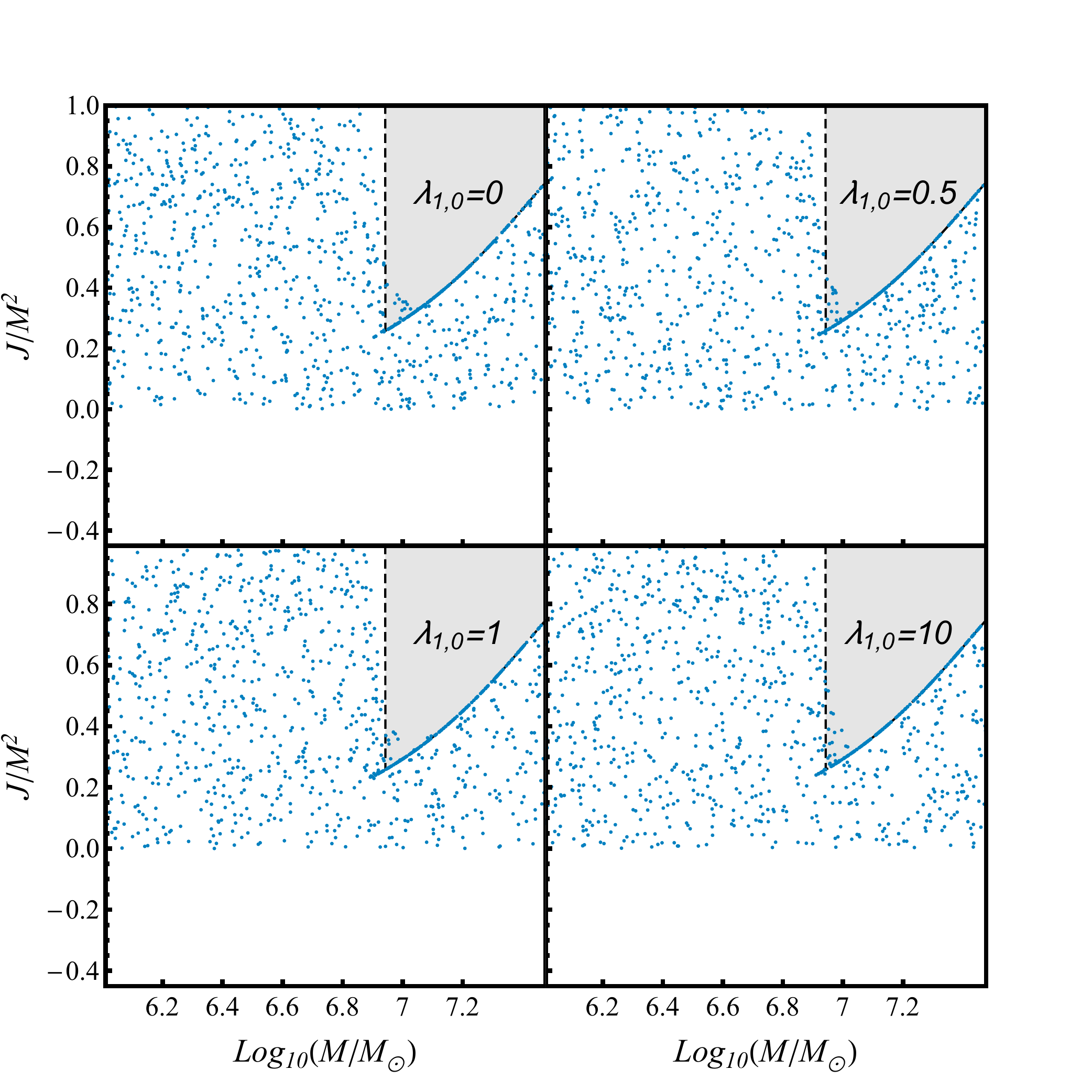}\label{fig:Regge1MS0m09}}
\subfloat[$M_{S0}=0.025M_{0}$]{\includegraphics[width=0.33\textwidth,clip]{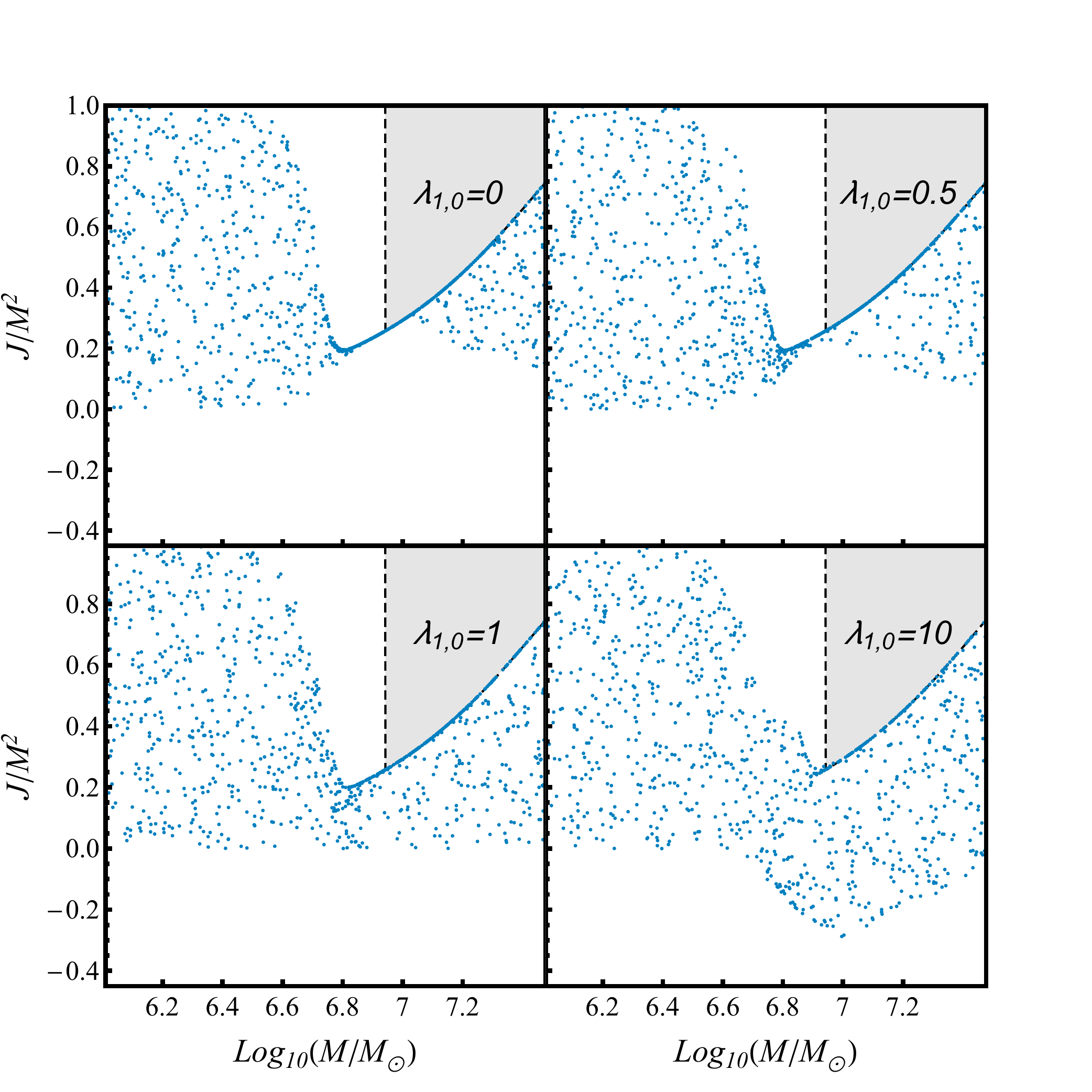}\label{fig:Regge1MS00025}}
\subfloat[$M_{S0}=0.05M_{0}$]{\includegraphics[width=0.33\textwidth,clip]{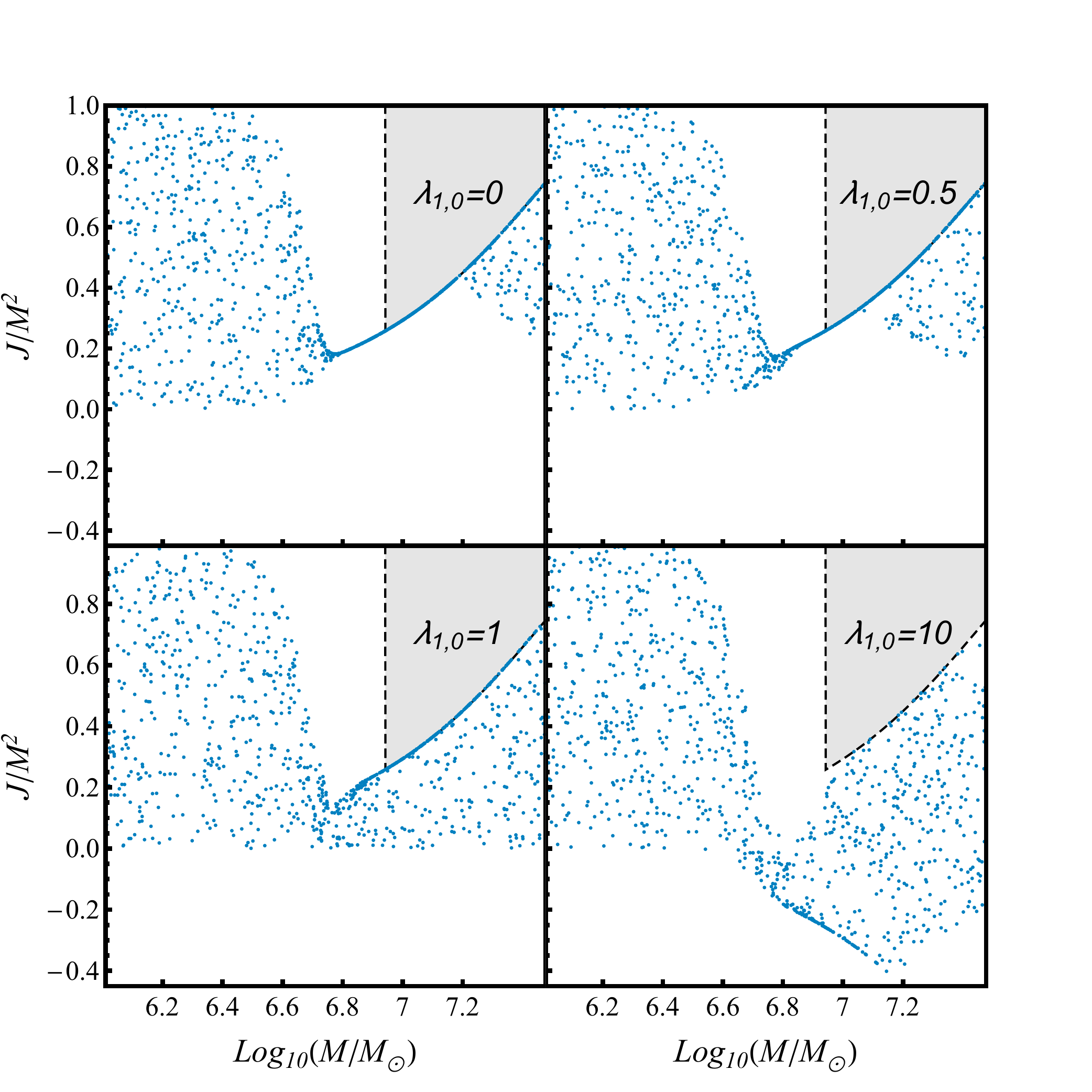}\label{fig:Regge1MS0005}}
\caption{\label{fig:Regge1}
BH Regge plane~\cite{Arvanitaki:2010sy} obtained from the (adiabatic) evolution of Model~I, i.e.,
a scalar field with mass $m_B=10^{-18}\,{\rm eV}$ and containing both co- and counterrotating, $l=1,m=\pm1$, modes.
We considered scalar cloud masses of~\protect\subref{fig:Regge1MS0m09} $M_{S0}=10^{-9}M_{0}$,
~\protect\subref{fig:Regge1MS00025} $M_{S0}=0.025M_{0}$,
and~\protect\subref{fig:Regge1MS0005} $M_{S0}=0.05M_{0}$,
and different initial relative amplitudes $\lambda_{1,0}$.
At $t=0$, we draw the initial BH mass and spin from a random distribution and then follow the evolution up to $t=10^8\,{\rm yr}$. 
Each point represents the final BH mass and spin. The shaded area is the Regge gap~\cite{Arvanitaki:2010sy} of the single, 
$l=m=1$ superradiant instability, i.e., $\lambda_{1,0}=0$ as computed in 
Ref.~\cite{Brito:2014wla}.
}
\end{figure*}

\begin{figure*}[th]
\centering
\subfloat[$M_{S0}=10^{-9}M_{0}$]{\includegraphics[width=.45\textwidth]{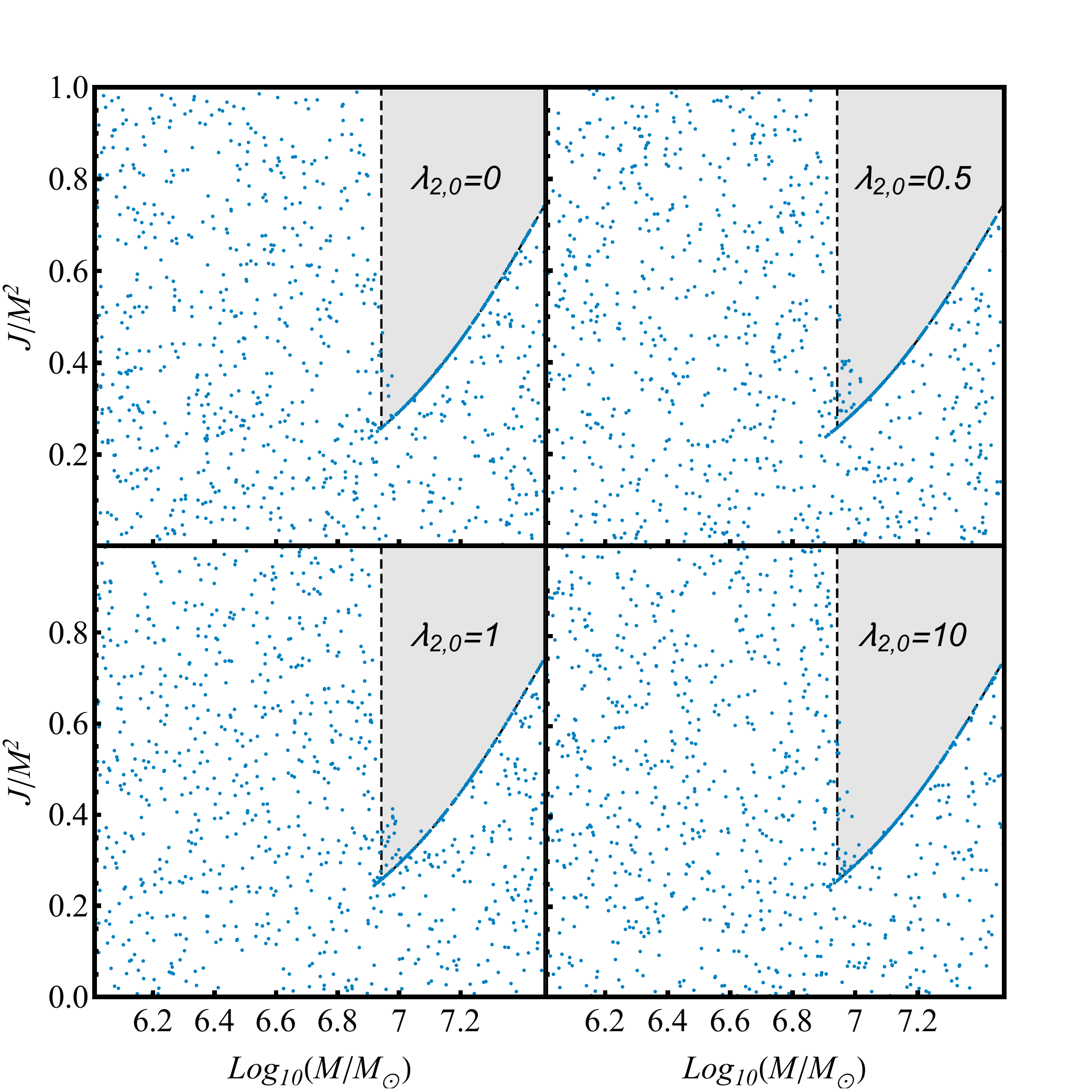}\label{fig:Regge2MS0m09}}
\subfloat[$M_{S0}=0.025M_{0}$]{\includegraphics[width=.45\textwidth]{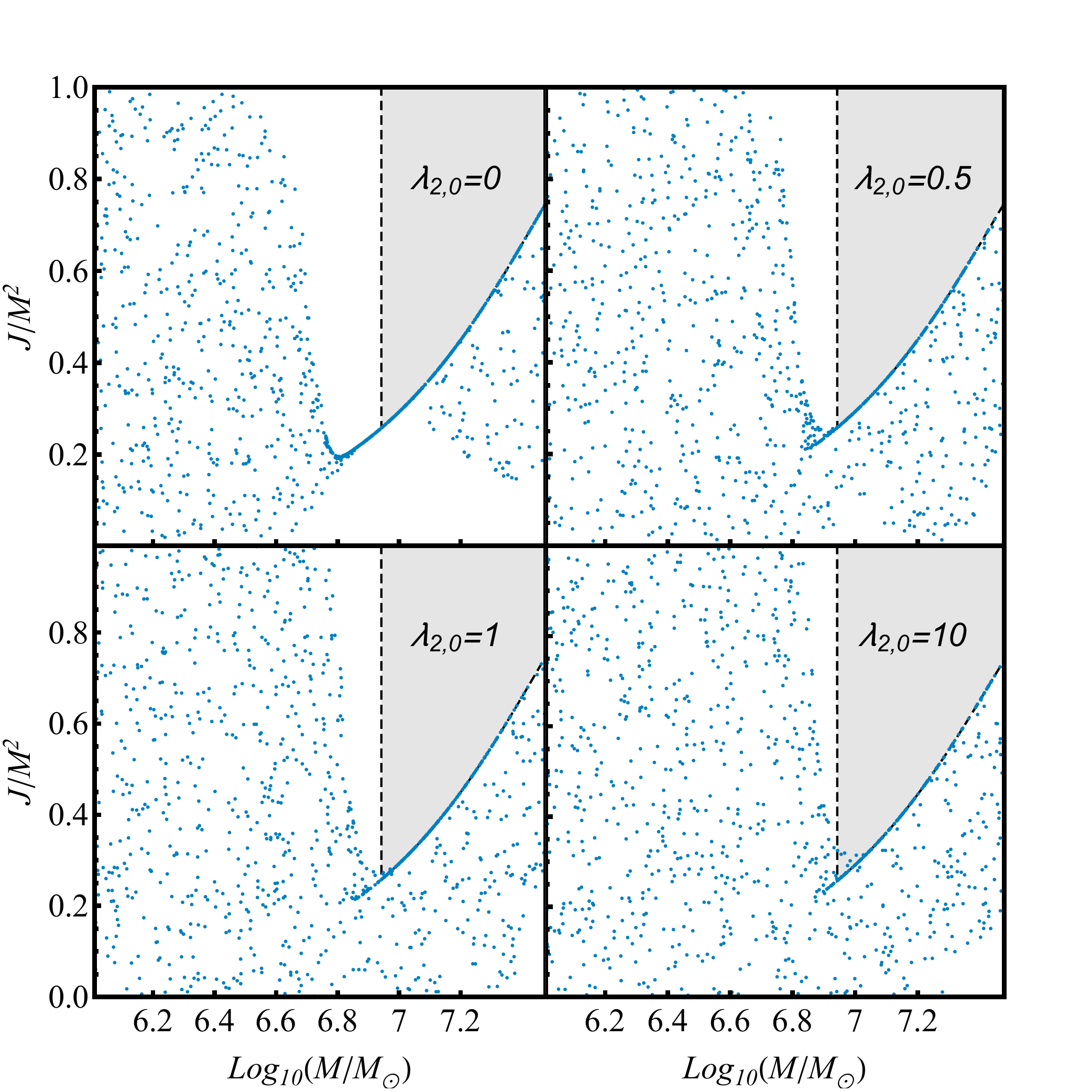}\label{fig:Regge2MS00025}}
\caption{\label{fig:Regge2}
Same as Fig.~\ref{fig:Regge1} but for Model~II and~\protect\subref{fig:Regge2MS0m09} Case~A 
and~\protect\subref{fig:Regge2MS00025} Case~B.
We evolved the systems for $t_{F} = 10^{8}$yr and different values of the initial relative 
amplitude $\lambda_{2,0}$.
The grey shaded areas denote the Regge gap due to a single, $l=m=1$ mode.
}
\end{figure*}
\begin{figure*}[th]
\centering
\subfloat[$M_{S0}=10^{-9}M_{0}$]{\includegraphics[width=.45\textwidth]{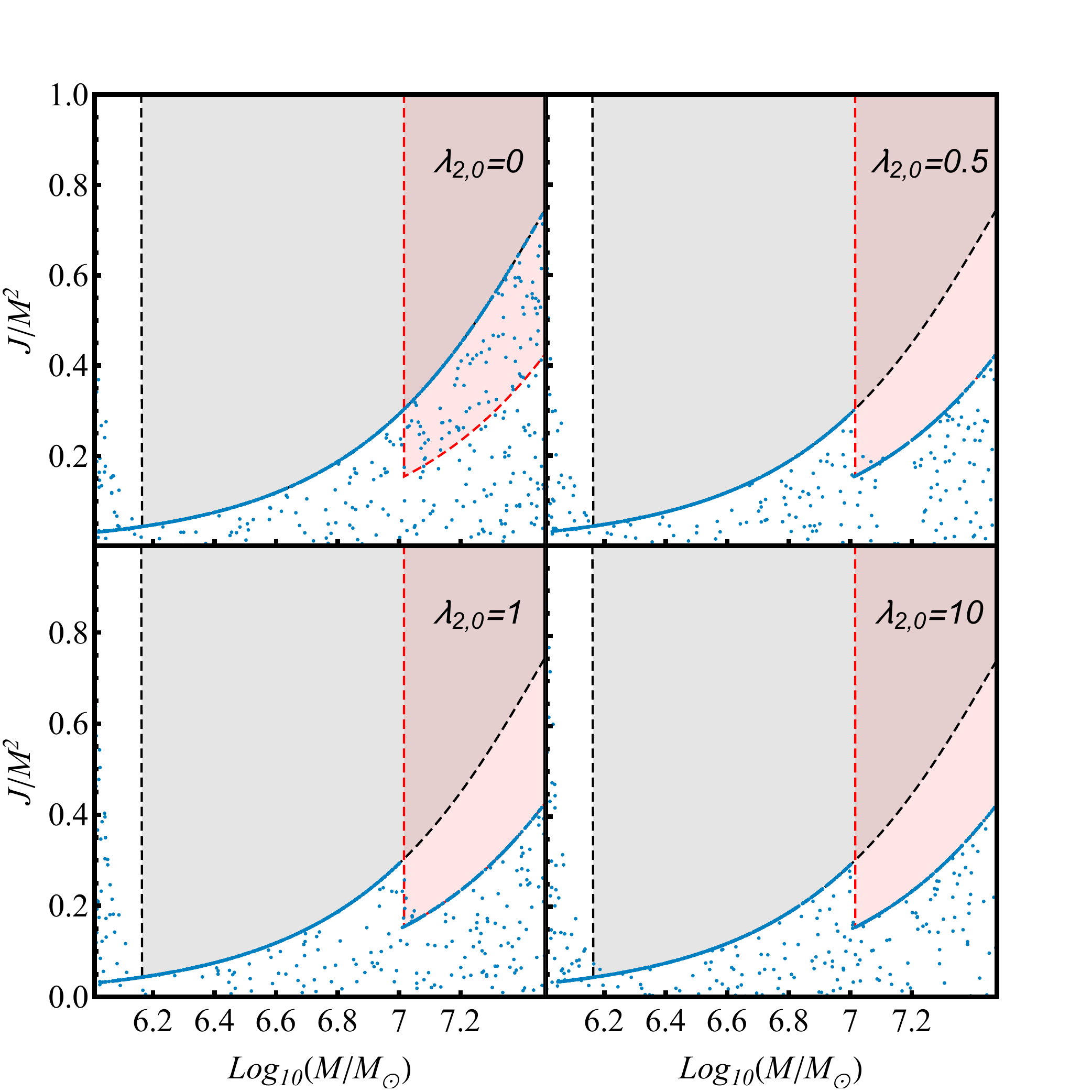}\label{fig:Regge2LongMS0m09}}
\subfloat[$M_{S0}=0.025M_{0}$]{\includegraphics[width=.45\textwidth]{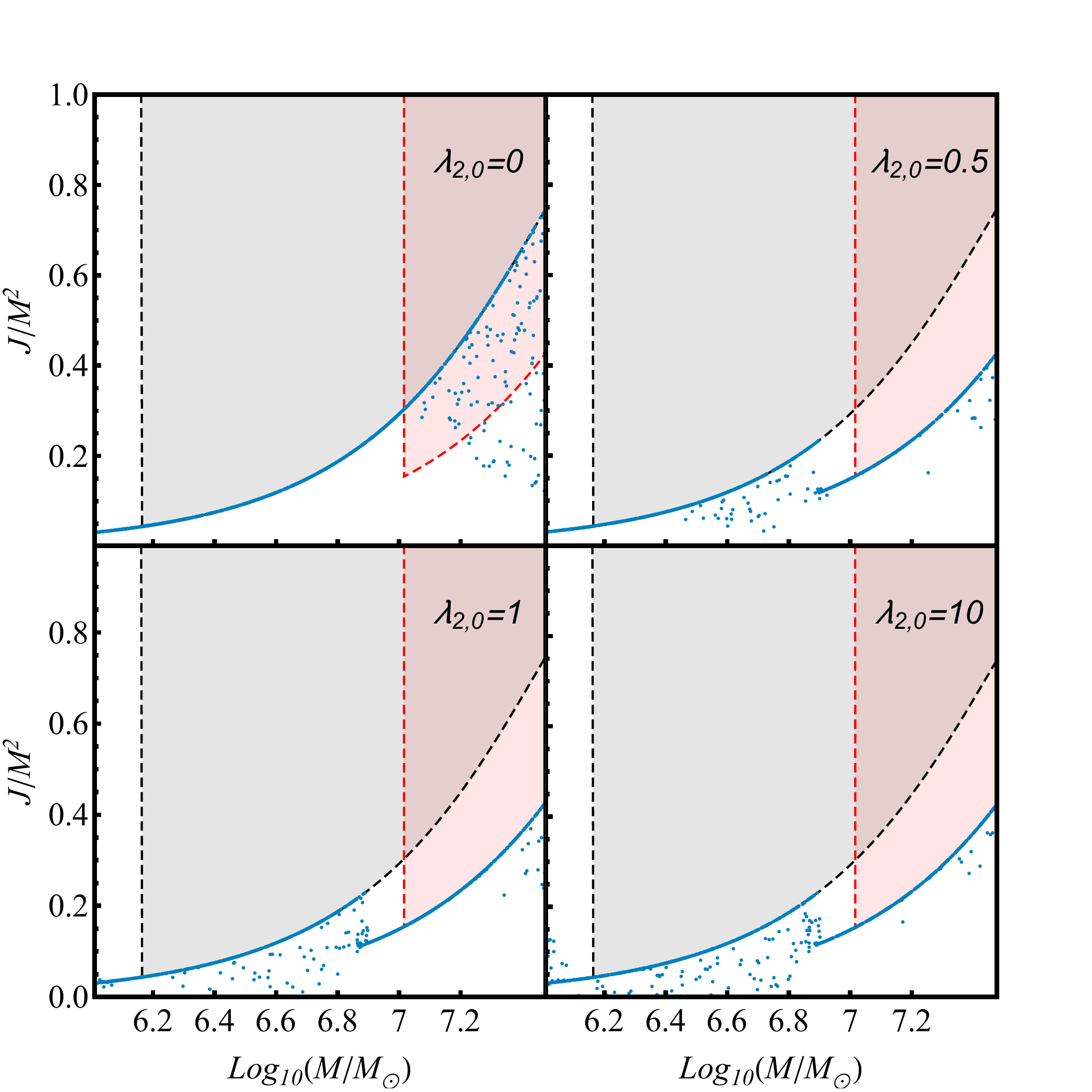}\label{fig:Regge2LongMS00025}}
\caption{\label{fig:Regge2Long}
Same as Fig.~\ref{fig:Regge1} but for Model~II and~\protect\subref{fig:Regge2MS0m09} Case~A 
and~\protect\subref{fig:Regge2MS00025} Case~B.
We evolved the systems for $t_{F} = 10^{15}$yr and different initial relative amplitudes 
$\lambda_{2,0}$.
The grey and red shaded areas denote the Regge gap due to a single, $l=m=1$ or $l=m=2$ 
mode, respectively.
}
\end{figure*}

Let us now focus on the mass-spin phase-space of the final BH encapsulated in its Regge-plot.
To identify it we performed a set of quasi-adiabatic evolutions whose results are shown in
Figs.~\ref{fig:Regge1},~\ref{fig:Regge2} and~\ref{fig:Regge2Long} for Models~I and~II, respectively.
In particular, we considered $m_B=10^{-18}\,{\rm eV}$ and $1000$ configurations starting at $t=0$ with 
a random distribution of the initial BH spin in the range $\chi_0\in(0,0.998)$ 
and masses in the range $\log_{10}M_0\in(6,7.5)$, so that the gravitational coupling 
$M_0\mu\in(0.0075,0.24)$. 

For comparison, we show the $\lambda_{i,0}=0$ case in the top-left panel of each plot in 
Figs.~\ref{fig:Regge1}--~\ref{fig:Regge2Long}.
Then, the final BH configuration avoids a specific region of the Regge plane, customarily 
dubbed Regge ``gap''~\cite{Arvanitaki:2010sy}. 
For a single $(l,m)$ mode, the shape of this gap is approximately given by~\cite{Brito:2014wla}
\begin{align}
\label{region}
\frac{J}{M^2}\gtrsim & \, \chi_{\rm crit}
\quad  \cup \quad 
M \gtrsim M_c 
\,.
\end{align}
where the critical spin $\chi_{\rm crit}$ is given in Eq.~\eqref{eq:ACritical},
$M_c$ is the value of $M$ that minimizes the spin when $\tau_{lm}=t_F$. An approximate formula is $M_c=\left({\frac{l+1}{C_l \mu ^{4l+6} t_F}}\right)^{\frac{1}{4l+5}}$~\cite{Brito:2014wla}.

{\noindent{\textbf{Model~I:}}}
We first focus on Model~I whose Regge planes are shown in Fig.~\ref{fig:Regge1}
for initial scalar cloud masses $M_{S0}=10^{-9}, 0.025, 0.05M_{0}$ and different relative 
amplitudes $\lambda_{1,0}$.
For each initial configuration, we followed the evolution of the system 
up to $t=t_F=10^8\,{\rm yr}\gg \tau_{11}$.

The evolution of a system containing only small scalar fluctuations $M_{S0}=10^{-9}M_{0}$, depicted in Fig.~\ref{fig:Regge1MS0m09},
is largely independent of the presence of a counter-rotating mode. In particular, it exhibits
the same exclusion regions in the Regge plane as those induced by the $l=m=1$ superradiant evolution~\cite{Brito:2014wla}.

If, instead, the scalar cloud already stores a significant fraction of the BH's mass -- of the order of a few percent --
the spin--mass phase-space of the final BH exhibits more structure; see Figs.~\ref{fig:Regge1MS00025} and~\ref{fig:Regge1MS0005}.
We identify three main features:
\begin{enumerate*}[label={(\roman*)}]
\item we still find gaps in the Regge plane consistent with those of the standard 
superradiant evolution. However, their onset occurs for smaller masses as the scalar cloud 
mass increases, even in the single-mode case.
This can be explained considering that a bigger initial value of 
the scalar cloud mass implies a larger energy flux rate via 
Eq.~\eqref{eq:EdotS1-1omegaIpiccolo} and, consequently, a shorter instability time scale.
Furthermore, if $\lambda_{1,0}\gtrsim10$, we start populating the low-mass end of the Regge gap. This is not surprising
as the absorption of (counter-rotating) modes decreases the BH spin while increasing its mass;
see e.g. Fig.~\ref{fig:M1largeA};
\item we find additional gaps in the Regge plane, just below the superradiant threshold, if
the scalar cloud is dominated by the $m=+1$ mode; see top panels of Figs.~\ref{fig:Regge1MS00025} and~\ref{fig:Regge1MS0005}.
BHs that are below the threshold will absorb part of the predominantly co-rotating cloud whose mass is 
$\mathcal{O}(1)\%$ of the BH mass. Hence their mass and spin will increase towards the superradiant threshold.
Should they supercede it
the superradiant instability will become active and drive the system towards the threshold from
above. That is to say, that the superradiant threshold appears to be an attractor if the initial scalar cloud mass is sufficiently
large and dominated by potentially superradiant (i.e. $m=+1$) modes;
\item if the initially large scalar cloud is instead dominated by counter-rotating modes, i.e. $\lambda_{1,0}\gtrsim10$,
as shown in the right-bottom panels of Figs.~\ref{fig:Regge1MS00025} and~\ref{fig:Regge1MS0005},
these additional holes disappear. Instead we find BHs with a negative final spin relative 
to its initial one.
This can be understood as follows:
BHs that have almost vanishing initial spin will absorb the counter-rotating modes that further decrease the BHs' spin.
Interestingly, one could now view the system containing a BH with negative spin and cloud with $m=-1$ modes
as co-rotating (but in the opposite direction as before), i.e., the situation is the 
same as that of a BH with positive spin surrounded by a $m=+1$ cloud. This, again, should 
suffer from the superradiant instability 
if the BH is spun up sufficiently.
Indeed, the bottom-right panel of Fig.~\ref{fig:Regge1MS0005} seems to exhibit such a new attractor line.
\end{enumerate*}

{\noindent{\textbf{Model~II:}}}
The Regge planes for Model~II are shown in Figs.~\ref{fig:Regge2} and~\ref{fig:Regge2Long} for an evolution time of 
$t_{F}=10^{8}\,{\rm yr}\gg\tau_{11}$ and $t_{F}=10^{15}\,{\rm yr}\gg\tau_{22}$, 
respectively. Although the latter time scale is larger than the age of the universe it 
allows us to explore features in the Regge plane due to both
the $l=m=1$ and $l=m=2$ instability for the supermassive BHs under consideration.
Note, furthermore, that this time scales with the BH mass. So for a much lighter, 
stellar-mass BH of $\mathcal{O}(10)M_{\odot}$
(and scalar of $m_{\rm B}\sim 10^{-12} {\rm eV}$)
we would observe features of Fig.~\ref{fig:Regge2Long} after $t_{F}\sim 10^{9}$yr.

The Regge plots of Fig.~\ref{fig:Regge2} only exhibit the $l=m=1$ Regge gap since we evolved the systems for times 
that are significantly shorter than the $l=m=2$ instability timescale.
If the scalar starts off as only a small fluctuation, i.e. Case~A,
the Regge gaps are identical to those of the single, $l=m=1$ mode.
That is, they are independent of the presence of a secondary mode as shown in Fig.~\ref{fig:Regge2MS0m09}.
The Regge gap itself is consistent with the estimate~\eqref{region}.

In Fig.~\ref{fig:Regge2MS00025} we consider a larger cloud with $M_{S0}=0.025M_{0}$.
As we already saw in Model~I, the onset of the superradiant instability is shifted towards smaller BH masses
as we increase the  scalar cloud mass.
In particular, the threshold $M_c$ is a complicate function of the parameters and differs 
from the expression below Eq.~\eqref{region}.
As before, we find an additional gap {\textit{below}} the superradiant threshold for sufficiently massive scalar clouds;
see top left panel of Fig.~\ref{fig:Regge2MS00025}.

Let us now turn our attention to the long-time evolution depicted in Fig.~\ref{fig:Regge2Long}
where we capture both the $l=m=1$ and $l=m=2$ phases.
For small initial seeds, depicted in Fig.~\ref{fig:Regge2LongMS0m09}, we observe the appearance of two Regge gaps
consistent, respectively, with the $l=m=1$ and $l=m=2$ superradiant evolution.
The details are independent of the (initial) relative amplitude $\lambda_{2,0}$ and appear
as soon as the secondary mode is switched on.

The Regge plane becomes more complex as we increase the scalar cloud's mass to a few percent of the
BH mass; see Fig.~\ref{fig:Regge2LongMS00025}.
In particular, we see the formation of a second gap below the superradiant threshold for $l=m=1$ if $\lambda_{2,0}=0$
and below the $l=m=2$ threshold as soon as $\lambda_{2,0}\neq0$.
Again, this can be understood as BHs that start outside the superradiant regime, but absorb mass (and angular momentum) 
from the scalar cloud until they reach the threshold.
Finally, the critical mass parametrizing the onset of the superradiant instability decreases, now for the $l=m=2$ case.

To summarize, even if a scalar condensate surrounding a BH contains counterrotating or higher multipole modes, 
in all cases studied here the holes in the Regge plane persist and yield a larger 
and more complex exclusion region.

\section{Discussion}~\label{sec:Discussion}
We have investigated the evolution of the BH superradiant instability against ultralight 
scalar fields with an initial configuration described by a superposition of modes. We 
focused on the case $M\mu\ll1$ which allows for a Newtonian description of the 
condensate and for a quasi-adiabatic approximation due to the separation of scales
between the instability time scale and the dynamical time scale of the BH.

Our analysis shows that the evolution of the superradiant instability in the presence of 
an initial superposition of modes is very rich and diverse.
The evolution of the system depends strongly on the energy of the scalar
seed and on the gravitational coupling $M\mu$. If the seed energy 
is a few percent of the BH mass, a BH surrounded by a mixture of superradiant and 
nonsuperradiant modes with comparable amplitudes might not even undergo a superradiant 
unstable phase, depending on the value of the boson mass.

Our analysis adds
to the numerical results of Refs.~\cite{Okawa:2014nda,Witek2018inprep},
where the authors explore the interplay between a highly spinning BH and massive scalars 
or vectors composed of multimode data and with $M\mu\sim\mathcal{O}(0.5)$. Indeed, our 
simple adiabatic approximation in the small-$M\mu$ limit is in remarkably good 
agreement with the evolution presented in Ref.~\cite{Okawa:2014nda}.
On the other hand, if the seed energy is much smaller than a few percent of the BH mass 
--~as in the most natural and likely scenario in which the instability is triggered by quantum 
fluctuations~-- the effect of 
nonsuperradiant modes is negligible. 

This implies that the only case in which the evolution of the superradiant instability is 
affected by multiple modes is when the BH is initially surrounded by a nonnegligible scalar 
environment, or if it is formed out of the coalescence of two BHs merging with their own 
scalar clouds. 
This latter scenario might be relevant only for a fraction of sources, in particular for 
massive BHs formed out of the merger of two BHs surrounded by their own condensates. In certain cases the time scale for 
GW dissipation of the condensate can exceed the age of the universe so a BH might form in a scalar-rich environment. 
This might be relevant for searches of ultralight fields with LISA~\cite{Brito:2017zvb,Brito:2017wnc}, since 
supermassive BHs are expected to form hierarchically.
In these cases the initial configuration of ultralight fields 
around BHs is generically a superposition of (superradiant and nonsuperradiant) modes, 
and the initial mass of the scalar configuration might be large enough to suppress the 
instability.
We leave a more detailed analysis of such a binary and event rate estimates for future work.

Likewise, the BH Regge plane is also affected by the presence of nonsuperradiant modes 
when the initial scalar mass is a sizeable fraction of the BH mass. The 
pattern of the Regge holes is more involved and additional forbidden regions can appear, 
depending on the parameters. 
Interestingly, the region forbidden in the single-mode case is also forbidden in the 
presence of nonsuperradiant modes, i.e. the original Regge holes are not populated even 
when the superradiant instability is absent. This is due the absorption of large 
counter-rotating modes which decrease the BH spin.

Our analysis can be extended in several directions. We have neglected mode mixing and 
possible transfer of energy between modes (e.g., turbulence) which might significantly 
change the overall picture. We have also neglected scalar self-interactions which --~if 
sufficiently strong~-- are known to quench the instability and give rise to interesting 
nonlinear effects such as ``bosenovas''~\cite{Yoshino:2012kn,Yoshino:2015nsa}. 
Likewise, we have neglected axion-like couplings to the electromagnetic 
field, which might also quench the instability through a different 
channel~\cite{Boskovic:2018lkj,Ikeda:2018nhb}.
We have also neglected accretion of 
ordinary matter; in light of the analysis of Ref.~\cite{Brito:2014wla}, we expect that 
including accretion should be a straightforward extension that would not give a 
substantial contribution to the understanding of the problem.
Furthermore, although we focused on scalar fields, it is likely that the qualitative 
features of the evolution will be the same also for massive vector (Proca) and massive 
tensor fields, as indicated by nonlinear simulations that will appear soon~\cite{Witek2018inprep}.


Finally, a natural extension of our work is to investigate whether the presence of 
multiple modes can also suppress the ergoregion instability of BH 
mimickers~\cite{Cardoso:2007az,Cardoso:2008kj,Pani:2010jz,Maggio:2017ivp,Maggio:2018ivz}, 
since the latter shares~\cite{Superradiance} many features with the superradiant 
instability discussed here.

\begin{acknowledgments}
We are indebted to Emanuele Berti, Roberto Emparan, and Vitor Cardoso for important 
comments on a first draft version of this work and to Richard Brito for relevant comments 
on the the current draft.
G.F. wishes to thank King's College London for financial support and the Royal Society for the PhD studentship provided under Research Grant RGF\textbackslash R1\textbackslash 180073.
P.P. acknowledges financial support provided under the European Union's H2020 ERC,  
Starting Grant agreement no.~DarkGRA--757480 and support from the Amaldi 
Research Center funded by the 
MIUR program ``Dipartimento di Eccellenza''~(CUP: B81I18001170001).
H.W. was supported by the European Union's H2020 research and innovation program under Marie Sklodowska-Curie grant agreement 
No.~{\textit{BHstabNL-655360}}
and acknowledges financial support provided by the Royal Society University Research Fellowship {\textit{UF160547}}
and Royal Society Research Grant RGF\textbackslash R1\textbackslash 180073.
The authors would like to acknowledge networking support by the COST Action CA16104.
H.W. thanks the Yukawa Institute for Theoretical Physics at
Kyoto University for their hospitality during the workshop YITP-T-17-02 on ``Gravity and Cosmology 2018'' and the
YKIS2018a symposium on ``General Relativity --~The Next Generation''.
%
We thankfully acknowledge the computer resources at Marenostrum IV, Finis Terrae II and LaPalma and the technical
support provided by the Barcelona Supercomputing Center via the PRACE grant Tier-0 PPFPWG, and via the BSC/RES grants
AECT-2017-2-0011, AECT-2017-3-0009 and AECT-2018-1-0014.
The authors thankfully acknowledge the computer resources and the technical support provided by the PRACE Grant No.\ 2018194669 ``FunPhysGW: Fundamental Physics in the era of gravitational waves'', STFC DiRAC Grant No.\ ACTP186 ``Extreme Gravity and
Gravitational Waves'' and STFC DiRAC Grant No.\ ACSP191 ``Exploring fundamental fields with strong gravity''. 
\end{acknowledgments}
%

\appendix

\section{GW emission from the scalar condensate} \label{app:GWs}

Owing to the separation of scales between the size of the cloud and the BH size for 
$M\mu\ll 1$, the GW emission can be approximately analyzed taking the source to lie in a 
flat~\footnote{We note that the flat-spacetime approximation yields a different 
prefactor for the GW fluxes emitted from the cloud relative to the case in which the background 
spacetime is described by a Schwarzschild metric~\cite{Yoshino:2013ofa,Brito:2013yxa}. 
The difference between the two cases is small and we adopt here a flat-spacetime approximation for simplicity.} 
background~\cite{Yoshino:2013ofa}. Because the source is incoherent $1/\omega\ll r_{\rm 
cloud}$, the quadrupolar approximation fails. In the fully relativistic regime, the 
gravitational radiation generated is best described by the Teukolsky formalism for 
gravitational perturbations~\cite{Teukolsky:1973ha}.

\subsection{General two modes case: $(l,m)$ and $(l',m')$}
The gravitational radiation is described by the Newman-Penrose scalar $\psi_4$, which, in the flat spacetime approximation, can be decomposed as
\begin{equation}
	\psi_4\tonde{r,t,\theta,\phi}=\sum_{j=0}^{\infty}\sum_{k=-j}^{j}\int_{-\infty}^{+\infty}{\dd\omega\frac{R_{jk\omega}(r)}{r^4}\,_{-2}Y_{jk}\tonde{\theta,\phi}}\,e^{-i\omega t}\,, \label{eq:NewmanPenroseScalar}
\end{equation}
where the radial function $R_{jk\omega}(r)$ satisfies the inhomogeneous Teukolsky equation,
\begin{align}
\label{eq:InhomogeneousRadialTeukolsky}
	&r^2R_{jk\omega}''-2(r-M)R_{jk\omega}' +
\\
	&\quadre{\omega^2r^2-4i\omega(r-3M)-(j+1)(j+2)}R_{jk\omega}=-T_{jk\omega} 
\,.\nonumber
\end{align}
The source term $T_{jk\omega}$ is given by~\cite{Poisson:1993vp}
\begin{align}
	\frac{T_{jk\omega}}{2\pi}=&2\quadre{\tonde{j-1}j\tonde{j+1}\tonde{j+2}}^{1/2}r^4\,_0T \nonumber \\
	&+2\quadre{2\tonde{j-1}\tonde{j+2}}^{1/2}r^2\mathcal{L}\tonde{r^3\,_{-1}T} \nonumber \\
	&+r\mathcal{L}\quadre{r^4\,\mathcal{L}\tonde{r\,_{-2}T}}, \label{eq:SourceTermEquation}
\end{align}
where we have defined $\mathcal{L}\equiv\partial_r+i\omega$ and
\begin{equation}
	_ST\equiv\frac{1}{2\pi}\int{\dd\Omega\,\dd t\,\Theta_S \,_S\bar{Y}_{jk}\,e^{i\omega t}},\label{eq:Integrals}
\end{equation}
where $\Theta_S=\covar{T}{nn},\,\covar{T}{n\bar{m}},\covar{T}{\bar{m}\bar{m}}$ for $S=0,\,-1,\,-2$, respectively.

The source term $T_{jk\omega}$ is related to the scalar field stress-energy tensor $\covar{T}{\mu\nu}$ through the tetrad projections
 \begin{align}
 	\covar{T}{nn}&=\covar{T}{\mu\nu}\,n^\mu n^\nu, \\
 	\covar{T}{n\bar{m}}&\equiv\covar{T}{\mu\nu}\,n^\mu\bar{m}^\nu, \\
 	\covar{T}{\bar{m}\bar{m}}&\equiv\covar{T}{\mu\nu}\,\bar{m}^\mu\bar{m}^\nu,
 \end{align}
where
\begin{align}
	n^\mu &\equiv\frac{1}{2}\tonde{1,-1,0,0},\\
	\bar{m}^\mu &\equiv\frac{1}{\sqrt{2}r}\tonde{0,0,1,-\frac{i}{\sin\vartheta}}.
\end{align}

For a scalar configuration with two modes with $(l,m)$ and $(l',m')$, the contributions to the source term are given by a sum over several active modes $(j,k)$, defined by the non vanishing contributions of the integrals~\eqref{eq:Integrals} that are strictly dependent on the values of $(l,m)$ and $(l',m')$. The contributions will feature two frequencies $\omega=\pm2\omega_R$, due to the fact that $T_{jk\omega}$, computed through Eq.~\eqref{eq:SourceTermEquation}, contains only terms $\propto\delta\tonde{\omega\pm2\omega_R}$.\

Once the source term is known, the radial equation~\eqref{eq:InhomogeneousRadialTeukolsky} can be solved using the Green's function. The latter can be found by considering two linearly independent solutions of the homogeneous equation associated with Eq.~\eqref{eq:InhomogeneousRadialTeukolsky}, with the following asymptotic behavior~\cite{Sasaki:2003xr},
\begin{equation}
	R^H\rightarrow
	\begin{cases}
		r^4\,e^{-ikr}\qquad\qquad\qquad\qquad\quad\,\quad r\rightarrow0, \\
		r^3\,B_{out}\,e^{i\omega r}+r^{-1}\,B_{in}\,e^{-i\omega r}\quad\quad r\rightarrow\infty,
	\end{cases} \label{eq:Rh}
\end{equation}
\begin{equation}
	R^\infty\rightarrow
	\begin{cases}
		A_{out}\,e^{ikr}+r^4\,A_{in}\,e^{-ikr}\,\,\qquad\,\quad r\rightarrow0, \\
		r^3\,e^{i\omega r}\qquad\qquad\qquad\qquad\,\,\quad\quad r\rightarrow\infty,
	\end{cases} \label{eq:Rinfty}
\end{equation}
where $k=\omega-m\Omega_H$, $\graffe{A,B}_{in,out}$ are constants. Owing to the flat spacetime approximation, the tortoise coordinate usually defined to deal with these kind of problems coincides with the standard radial coordinate.

Imposing ingoing boundary conditions at the horizon and outgoing boundary conditions at infinity, one finds that the solution of Eq.~\eqref{eq:InhomogeneousRadialTeukolsky} is given by~\cite{Sasaki:2003xr}
\begin{align}
	R_{jk\omega}(r)=\frac{R^\infty}{W} \int_0^r{\dd r'}\frac{R^H\,T_{jk\omega}}{r^4}+\frac{R^H}{W} \int_r^\infty{\dd r'}\frac{R^\infty\,T_{jk\omega}}{r^4},
\end{align}
where $W=\tonde{R^\infty\partial_r R^H-R^H\partial_rR^\infty}/r=2i\omega B_{in}$ is the Wronskian, which is a constant by virtue of the homogeneous Teukolsky equation.
From the asymptotic solution of Eq.~\eqref{eq:InhomogeneousRadialTeukolsky} we find
\begin{equation}
	B_{in}=-\frac{C_1}{8\omega^2}\tonde{j-1}j\tonde{j+1}\tonde{j+2}e^{i(j+1)\frac{\pi}{2}},
\end{equation}
where $C_1$ is an arbitrary constant that we set to unity without loss of generality. 
The solution $R^H$ can be found through
\begin{equation}
	R^H=r^2\mathcal{L}\tonde{\mathcal{L}r\psi^H},
\end{equation}
where $\psi^H$ is the Regge-Wheeler function that at small frequencies reads	
\begin{equation}
	\psi^H\sim\omega rj_j\tonde{\omega r},
\end{equation}
where $j_j$ are the spherical Bessel functions of the first kind.
At radial infinity the solutions reads
\begin{equation}
	R_{jk\omega}\tonde{r\rightarrow\infty}\rightarrow\frac{R^\infty}{W}\int_0^r{\dd r'}\frac{R^H\,T_{jk\omega}}{r^4}\equiv\tilde{Z}_{jk\omega}^\infty\,r^3\,e^{i\omega r}.
\end{equation}

Since the frequency spectrum of the source $T_{jk\omega}$ is discrete with frequencies $\omega=\pm2\omega_R$, $\tilde{Z}_{jk\omega}^\infty$ can be written as
\begin{equation}
	\tilde{Z}_{jk\omega}^\infty=\sum_{q=1}^{2}Z^\infty_{jkq}\,\delta\tonde{\omega-\omega_q}, 
\end{equation}
where $\omega_1=2\omega_R$ and $\omega_2=-2\omega_R$.
Replacing the above equation in Eq.~\eqref{eq:NewmanPenroseScalar}, we obtain $\psi_4$ at radial infinity,
\begin{equation}
	\psi_4=\frac{1}{r}\sum_{j=0}^{\infty}\sum_{k=-j}^{j}\sum_{q=1}^{2}Z^\infty_{jkq}\,_{-2}Y_{jk}\tonde{\theta,\phi}\,e^{i\omega_q(r-t)}\,, \label{eq:Psi4InftyGeneral}
\end{equation} 
which can be written as
\begin{equation}
	\psi_4=\frac{1}{2}\tonde{\ddot{h}_+-i\ddot{h}_{\times}}, \label{eq:Psi4Polarization}
\end{equation}
where $h_+$ and $h_{\times}$ are the two independent GW polarizations.\ Then, using Eq.~\eqref{eq:Psi4InftyGeneral} in the previous relation and integrating twice with respect to the time, we obtain the gravitational waveform,
\begin{equation}
	h_+-ih_\times=\frac{2}{r}\sum_{j=0}^{\infty}\sum_{k=-j}^{j}\sum_{q=1}^{2}\frac{Z^\infty_{jkq}}{\omega^2_q}\,_{-2}Y_{jk}\tonde{\theta,\phi}\,e^{i\omega_q(r-t)}\,.
\end{equation} 

The energy flux carried by these waves at infinity is given by~\cite{Teukolsky:1974yv}
\begin{equation}
	\frac{d^2E}{dt\,d\Omega}=\lim_{r\to\infty}\sum_{q=1}^{2}\frac{r^2}{4\pi\omega_q^2}\modulo{\psi_4}^2\equiv\lim_{r\to\infty}\frac{r^2}{16\pi}\tonde{\dot{h}^2_++\dot{h}^2_\times}. \label{eq:GWEnergyFlux}
\end{equation}

Finally, combining the last two equations, we get the energy and angular momentum fluxes at radial infinity~\cite{Hughes:2001jr}
\begin{align}
	\frac{dE}{dt}\equiv\dot{E}_{GW}&=\sum_{j=0}^{\infty}\sum_{k=-j}^{j}\sum_{q=1}^{2}\frac{1}{4\pi\omega_q^2}\modulo{Z^\infty_{jkq}}^2\,, \label{eq:GWEnergyFluxGeneral} \\
	\frac{dJ}{dt}\equiv\dot{J}_{GW}&=\sum_{j=0}^{\infty}\sum_{k=-j}^{j}\sum_{q=1}^{2}\frac{k}{4\pi\omega_q^3}\modulo{Z^\infty_{jkq}}^2\,. \label{eq:GWAngularMomentumFluxGeneral}
\end{align}

\subsection{Particular cases}
We shall now apply the above results to the three models presented in the main text.

\paragraph{Model~I}.\ For the scalar
configuration~\eqref{eq:Psi1-1} the contributions to the source term are given by $(j=2,k=\pm2)$ with two frequencies $\omega=\pm2\omega_R$, since $T_{jk\omega}$ contains only terms $\propto\delta\tonde{\omega\pm2\omega_R}$. Then, the right-hand sides of Eqs.~\eqref{eq:GWEnergyFluxGeneral} and~\eqref{eq:GWAngularMomentumFluxGeneral} take the form
\begin{align}
	\sum_{j=0}^{\infty}\sum_{k=-j}^{j}\sum_{q=1}^{2}\frac{\modulo{Z^\infty_{jkq}}^2}{4\pi\omega_q^2}&=& \sum_{q=1}^{2}\frac{\modulo{Z^\infty_{22q}}^2+\modulo{Z^\infty_{2-2q}}^2}{4\pi\omega_q^2}\,, \\
	\sum_{j=0}^{\infty}\sum_{k=-j}^{j}\sum_{q=1}^{2}\frac{\modulo{Z^\infty_{jkq}}^2}{4\pi\omega_q^2}&=& \sum_{q=1}^{2}\frac{\modulo{Z^\infty_{22q}}^2-\modulo{Z^\infty_{2-2q}}^2}{2\pi\omega_q^3}\,.
\end{align}
In this particular case, using Eq.~\eqref{eq:Psi1-1} and considering the small $M\mu$-limit, we obtain
\begin{align}
	\dot{E}_{GW}&=\frac{32\pi^2}{5}\tonde{A_{11}^4+A_{1-1}^4}\tonde{M\mu}^6\,, \nonumber \\
	\dot{J}_{GW}&=\frac{1}{\omega_R}\frac{32\pi^2}{5}\tonde{A_{11}^4-A_{1-1}^4}\tonde{M\mu}^6. \nonumber
\end{align}
Finally, using Eq.~\eqref{eq:Amplitude1-1} we get Eqs.~\eqref{eq:EdotGW1-1} and~\eqref{eq:JdotGW1-1}.

\paragraph{Model~II}.\ For the scalar
configuration~\eqref{eq:Psi1122} in Eq.~\eqref{eq:SourceTermEquation}, we have different contributions relative to $(j=2,k=\pm2)$, $(j=3,k=\pm3)$ and $(j=4,k=\pm4)$. Furthermore, the contributions with $k>0$ are $\propto\delta\tonde{\omega-2\omega_R}$, while those with $k<0$ are $\propto\delta\tonde{\omega+2\omega_R}$.\ In this case the right-hand sides of Eqs.~\eqref{eq:GWEnergyFluxGeneral} and~\eqref{eq:GWAngularMomentumFluxGeneral} become
\begin{align}
	\sum_{j=0}^{\infty}&\sum_{k=-j}^{j}\sum_{q=1}^{2}\frac{1}{4\pi\omega_q^2}\modulo{Z^\infty_{jkq}}^2= \nonumber \\ &\sum_{j=2}^{4}\frac{1}{4\pi\omega_1^2}\modulo{Z^\infty_{jj1}}^2+\frac{1}{4\pi\omega_2^2}\modulo{Z^\infty_{j-j2}}^2\,, \label{eq:GWEnergyl2} \\
	\sum_{j=0}^{\infty}&\sum_{k=-j}^{j}\sum_{q=1}^{2}\frac{1}{4\pi\omega_q^2}\modulo{Z^\infty_{jkq}}^2= \nonumber \\ &\sum_{j=2}^{4}\frac{j}{4\pi\omega_1^3}\modulo{Z^\infty_{jj1}}^2+\frac{(-j)}{4\pi\omega_2^3}\modulo{Z^\infty_{j-j2}}^2\,. \label{eq:GWAMl2}
\end{align}
Using Eqs.~\eqref{eq:Psi1122},~\eqref{eq:GWEnergyl2} and~\eqref{eq:GWAMl2} for $M\mu\ll 1$, we get
\begin{align}
	&\dot{E}_{GW}=\frac{32}{5}A_{11}^4\pi^2\tonde{M\mu}^6 + \nonumber \\&\frac{16384}{1701}A_{11}^2A_{22}^2\pi^2\tonde{M\mu}^8+\frac{2097152}{413343}A_{22}^4\tonde{M\mu}^{10} \,, \\
	&\dot{J}_{GW}=\frac{32}{5\,\omega_R}A_{11}^4\pi^2\tonde{M\mu}^6+ \nonumber \\ &\frac{8192}{567\,\omega_R}A_{11}^2A_{22}^2\pi^2\tonde{M\mu}^8+\frac{4194304}{413343\,\omega_R}A_{22}^4\tonde{M\mu}^{10}\,. \nonumber
\end{align}
Finally, using Eq.~\eqref{eq:Amplitude1122}, the above equations reduce to 
\begin{eqnarray}
\dot{E}_{GW}&= & C_E \tonde{\frac{M_S}{M}}^2\tonde{M\mu}^{14}\,, \\
\dot{J}_{GW}&= & C_J \tonde{\frac{M_S}{M}}^2\tonde{M\mu}^{14}\,,
\end{eqnarray}
with
\begin{eqnarray}
 C_E &=& \frac{413343+2560\lambda_2^2\tonde{M\mu}^2\tonde{243+128\lambda_2^2\tonde{M\mu}^2}}{66134880\tonde{1+81\lambda_2^2}^2}\,, \nonumber\\
 C_J &=& \frac{413343+1280\lambda_2^2\tonde{M\mu}^2\tonde{243+512\lambda_2^2\tonde{M\mu}^2}}{66134880\,\omega_R\tonde{1+81\lambda_2^2}^2}\,. \nonumber
\end{eqnarray}

The contributions $\propto\lambda_{2}$ in the numerator are subleading when $M\mu\ll 1$. This implies that, in the considered limit, $\dot{E}_{GW},\dot{J}_{GW}\propto (M_S/M)^2(M\mu)^{14}$.
In this limit, the GW energy and angular momentum fluxes are given by Eqs.~\eqref{eq:EdotGW1122} and~\eqref{eq:JdotGW1122}.

\paragraph{Model~III}.\
The stress-energy tensor corresponding to the scalar
configuration~\eqref{eq:Psi1122} in Eq.~\eqref{eq:SourceTermEquation} yields contributions corresponding to $(j=2,k=\pm2)$, $(j=3,k=\pm3)$ and $(j=4,k=\pm4)$. Again, the ones with $k>0$ are $\propto\delta\tonde{\omega-2\omega_R}$, while for $k<0$ they are $\propto\delta\tonde{\omega+2\omega_R}$. The analysis is the same as for Model~II above; the right-hand sides of Eqs.~\eqref{eq:GWEnergyFluxGeneral} and~\eqref{eq:GWAngularMomentumFluxGeneral} are given by Eqs.~\eqref{eq:GWEnergyl2} and~\eqref{eq:GWAMl2}.

Finally, considering Eqs.~\eqref{eq:Psi1121},~\eqref{eq:GWEnergyl2} and~\eqref{eq:GWAMl2} for $M\mu\ll 1$, we get:
\begin{align}
	\dot{E}_{GW}&=\frac{32}{5}A_{11}^4\pi^2\tonde{M\mu}^6+  \nonumber \\ &\frac{8192}{5103}A_{11}^2A_{21}^2\pi^2\tonde{M\mu}^8+\frac{524288}{2893401}A_{21}^4\tonde{M\mu}^{10} \,, \\ 
	\dot{J}_{GW}&=\frac{32}{5\omega_R}A_{11}^4\pi^2\tonde{M\mu}^6+  \nonumber \\ &\frac{4096}{1701\,\omega_R}A_{11}^2A_{21}^2\pi^2\tonde{M\mu}^8+\frac{1048576}{2893401\,\omega_R}A_{21}^4\tonde{M\mu}^{10}\,, \nonumber
\end{align}

Using Eq.~\eqref{eq:Amplitude1121}, the equations above can be written as
\begin{align}
\dot{E}_{GW}&= C_E \tonde{\frac{M_S}{M}}^2\tonde{M\mu}^{14} \,,  \\
\dot{J}_{GW}&= C_J \tonde{\frac{M_S}{M}}^2\tonde{M\mu}^{14}\,,
\end{align}
with
\begin{eqnarray}
 C_E &=&\frac{2893401+1280\lambda_3^2\tonde{M\mu}^2\tonde{567+64\lambda_3^2\tonde{M\mu}^2}}{28934010\tonde{4+81\lambda_3^2}^2}\,, \nonumber\\
 C_J &=& \frac{2893401+640\lambda_3^2\tonde{M\mu}^2\tonde{1701+256\lambda_3^2\tonde{M\mu}^2}}{28934010\,\omega_R\tonde{4+81\lambda_3^2}^2}\nonumber\,.
\end{eqnarray}

Once again, the contributions $\propto\lambda_{3}$ in the numerator are subleading when $M\mu\ll 1$ and can be neglected, finally obtaining Eqs.~\eqref{eq:EdotGW1121} and~\eqref{eq:JdotGW1121}.

\section{Scalar energy and angular momentum fluxes at the horizon} \label{app:SRflux}

In this section we compute the adiabatic time variation of the mass $M_S$ and angular momentum $L_S$ of the condensate due to the superradiant instability.

In the Newtonian approximation, the condensate mass is given by Eq.~\eqref{eq:ScalarCloudMass}, whereas the $z$-component of the angular momentum of the condensate reads
\begin{equation}
	L_S=\int{\dd r\,\dd\theta\,\dd\phi\,r^2\sin\theta\tonde{xT^{0y}-yT^{0x}}},\label{eq:ScalarCloudAngularMomentum}
\end{equation}
where the quantity $xT^{0y}-yT^{0x}$ has to be expressed in spherical coordinates.

In order to include an adiabatic time dependence, we make the substitution $A_{lm}\to A_{lm} e^{t/\tau_{lm}}$ in the expression~\eqref{eq:superposition} of the mode $\Psi$. Clearly, $\tau_{lm}>0$ in the superradiant phase, whereas $\tau_{lm}<0$ otherwise.

\subsection{Model~I: $l=1$ and $m=\pm1$}
We start analyzing the case of a scalar cloud described by Eq.~\eqref{eq:Psi1-1}. Including the time dependence, the expression of the scalar cloud reads
\begin{align}
	\Psi =A_{11}\,&e^{\omega_{11}t}\,g_1(r)\cos\tonde{\phi-\omega_R t}\sin\theta+ \nonumber \\ &A_{1-1}\,e^{\omega_{1-1}t}\,g_1(r)\cos\tonde{\phi+\omega_R t}\sin\theta, \label{eq:Psi1-1TimeDependent}
\end{align} 
where we recall that $\omega_{lm}\equiv\omega_I$ with a given value of ($l$, $m$).
Using Eq.~\eqref{eq:ScalarCloudMass}, we obtain
\begin{equation}
	M_S(t)=\frac{32\pi 
M}{\tonde{M\mu}^4}\tonde{A_{11}^2\,e^{2\omega_{11}t}+A^2_{1-1}\,e^{2\omega_{1-1}t}}.
\end{equation} 
Note that $\omega_I\ll\omega_R$ in the small-$M\mu$ limit; an important consequence of the 
latter is the absence of terms proportional to $A_{11}A_{1-1}$ in the above formula. 
Using Eq.~\eqref{eq:Amplitude1-1} we get
\begin{equation}	
M_S(t)=\frac{M_S(0)}{1+\lambda_1^2}\tonde{e^{2\omega_{11}t}+\lambda_1^2e^{
2\omega_{1-1}t} },\label{eq:M_S(t)1-1}
\end{equation}
where $M_S(0)$ is the value of $M_S$ at $t=0$. Then, from Eq.~\eqref{eq:ScalarEnergyFlux} 
we obtain
\begin{equation}	
\dot{E}_{S}=M_S(0)\frac{2}{1+\lambda_1^2}\tonde{\omega_{11}\,e^{2\omega_{11}t}+\omega_{1-1
}\,\lambda_1^2e^{2\omega_{1-1}t}}\,,
\end{equation}
which can be expressed as a function of $M_S(t)$, $\lambda_1$ and $\omega_{lm}$ by isolating $M_S(0)$ in Eq.~\eqref{eq:M_S(t)1-1} and replacing it in the last expression:
\begin{equation}
	\dot{E}_{S}=2\,M_S(t)\tonde{\frac{\omega_{11}\,e^{2\omega_{11}t}+\lambda_1^2\,\omega_{1-1}\,e^{2\omega_{1-1}t}}{e^{2\omega_{11}t}+\lambda_1^2e^{2\omega_{1-1}t}}}.\label{eq:ScalarEnergyFlux1-1}
\end{equation}

Likewise, using Eq.~\eqref{eq:ScalarCloudAngularMomentum} for the configuration Eq.~\eqref{eq:Psi1-1TimeDependent}, we obtain the angular momentum of the condensate,
\begin{equation}
	L_S=\frac{32\pi 
}{M^3\mu^5}\tonde{A_{11}^2\,e^{2\omega_{11}t}-A^2_{1-1}\,e^{2\omega_{1-1}t}}\,, 
\label{eq:JS}
\end{equation}
or, equivalently,
\begin{equation}
	\dot{J}_S=2\,\frac{M_S(t)}{\mu}\tonde{\frac{\omega_{11}\,e^{2\omega_{11}t}-\lambda_1^2\,\omega_{1-1}\,e^{2\omega_{1-1}t}}{e^{2\omega_{11}t}+\lambda_1^2e^{2\omega_{1-1}t}}} \,.\label{eq:ScalarAngularMomentumFlux1-1}
\end{equation} 
Note that by using Eqs.~\eqref{eq:JS} and~\eqref{eq:M_S(t)1-1} the
initial angular momentum of the cloud, $L_{S0}\equiv L_S(0)$, is fixed in terms of the 
the initial mass $M_{S0}\equiv M_S(0)$, the initial relative amplitude of the modes, 
$\lambda_{10}$, the initial BH parameters $M_0$ and $\chi_0$, and the gravitational 
coupling $M_0\mu$.

In the quasi-adiabatic approximation there should not be any explicit time dependence because all the quantities of interest implicitly vary over the time evolution of the system. To remove the explicit time dependence in Eqs.~\eqref{eq:ScalarEnergyFlux1-1} and~\eqref{eq:ScalarAngularMomentumFlux1-1}, we can consider an average over a time $2\pi/\omega_R\sim2\pi/\mu$, which is the characteristic orbital time period of the scalar condensate,
\begin{align}
	\media{\dot{E}_S}&=2M_S\int_{0}^{2\pi/\mu}{\dd t\,\frac{\omega_{11}\,e^{2\omega_{11}t}+\lambda_1^2\,\omega_{1-1}\,e^{2\omega_{1-1}t}}{e^{2\omega_{11}t}+\lambda_1^2e^{2\omega_{1-1}t}}}, \\
	\media{\dot{J}_S}&=2\frac{M_S}{\mu}\int_{0}^{2\pi/\mu}{\dd t\,\frac{\omega_{11}\,e^{2\omega_{11}t}-\lambda_1^2\,\omega_{1-1}\,e^{2\omega_{1-1}t}}{e^{2\omega_{11}t}+\lambda_1^2e^{2\omega_{1-1}t}}},
\end{align}
where $M_S$, $\omega_{11}$ are $\omega_{1-1}$ are treated as constants.
Performing the integrals we obtain
\begin{eqnarray}
\media{\dot{E}_S}&=&\frac{M_S\,\mu}{2\pi} \quadre{\ln\tonde{e^{\frac{4\pi\omega_{11}}{\mu}}+\lambda_1^2\,e^{\frac{4\pi\omega_{1-1}}{\mu}}}-\ln\tonde{1+\lambda_1^2}} \,, \nonumber\\
\media{\dot{J}_S}&= &  {\frac{{\tonde{\omega_{11}+\omega_{1-1}}\media{\dot{E}_S}}}{\mu\tonde{\omega_{11}-\omega_{1-1}}}-\frac{4 M_S\omega_{11}\omega_{1-1}}{\mu\tonde{\omega_{11}-\omega_{1-1}}}}\,.\nonumber
\end{eqnarray}
%
Finally, by expanding the above relations in the $M\mu\ll1$ limit, we obtain Eqs.~\eqref{eq:EdotS1-1omegaIpiccolo}--\eqref{eq:JdotS1-1omegaIpiccolo}.
Note that the final result would be the same if the averages were performed over several orbital periods.

\subsection{Model~II: $l=m=2$ and $l=m=1$}
We now consider the case of a scalar cloud described by Eq.~\eqref{eq:Psi1122}. After including the time dependence, the expression of the scalar condensate reads
\begin{align}
	\Psi=A_{11}\,&e^{\omega_{11}t}\,g_1(r)\cos\tonde{\phi-\omega_R t}\sin\theta+ \nonumber \\ &A_{22}\,e^{\omega_{22}t}\,g_2(r)\cos\tonde{2\phi-\omega_Rt}\sin^2\theta\,. \label{eq:Psi1122TimeDependent}
\end{align} 
Using Eq.~\eqref{eq:ScalarCloudAngularMomentum} for this configuration, we get
\begin{equation}
	L_S=\frac{32\pi 
}{M^3\mu^5}\tonde{A_{11}^2\,e^{2\omega_{11}t}+162A_{22}^2\,e^{2\omega_{1-1}t}}\,, 
\label{eq:LSmodelII}
\end{equation}
and the same computation described above for Model~I yields
\begin{eqnarray}
\dot{E}_{S}&=&2\,M_S(t)\tonde{\frac{\omega_{11}\,e^{2\omega_{11}t}+81\,\lambda_2^2\,\omega_{22}\,e^{2\omega_{22}t}}{e^{2\omega_{11}t}+81\,\lambda_2^2e^{2\omega_{22}t}}}\,,\nonumber\\
\dot{J}_S&=&2\,\frac{M_S(t)}{\mu}\tonde{\frac{\omega_{11}\,e^{2\omega_{11}t}+162\,\lambda_2^2\,\omega_{22}\,e^{2\omega_{22}t}}{e^{2\omega_{11}t}+81\,\lambda_2^2e^{2\omega_{22}t}}}\,.\nonumber	
\end{eqnarray}
In this case, the time average gives
\begin{widetext}
 \begin{eqnarray}
		\media{\dot{E}_S}&=&\frac{M_S\,\mu}{2\pi}\quadre{\ln\tonde{e^{\frac{4\pi\omega_{11}}{\mu}}+81\,\lambda_2^2\,e^{\frac{4\pi\omega_{22}}{\mu}}}-\ln\tonde{1+81\,\lambda_2^2}}\,, \\
		\media{\dot{J}_S}&= & M_S\frac{\graffe{4\pi\omega_{11}\omega_{22}+\tonde{\omega_{11}-2\omega_{22}}\mu\quadre{\ln\tonde{e^{\frac{4\pi\omega_{11}}{\mu}}+81\,\lambda_2^2\,e^{\frac{4\pi\omega_{22}}{\mu}}}-\ln\tonde{1+81\,\lambda_2^2}}}}{2\pi\mu\tonde{\omega_{11}-\omega_{22}}}.
\end{eqnarray}
\end{widetext}
Finally, by expanding the above relations in the $M\mu\ll1$ limit, we obtain Eqs.~\eqref{eq:EdotS1122omegaIpiccolo}--\eqref{eq:JdotS1122omegaIpiccolo}.
\subsection{Model~III: $l=1,2$ with $m=1$}
At last let us consider the case of a scalar cloud described by Eq.~\eqref{eq:Psi1121}. In this case the scalar condensate reads
\begin{align}
	\Psi=A_{11}\,&e^{\omega_{11}t}\,g_1(r)\cos\tonde{\phi-\omega_R t}\sin\theta+ \nonumber \\ & A_{21}\,e^{\omega_{21}t}\,g_2(r)\cos\tonde{\phi-\omega_Rt}\cos\theta\sin\theta\,,\label{eq:Psi1121TimeDependent}
\end{align}
and we obtain
\begin{eqnarray}
	\dot{E}_{S}&=&2\,M_S(t)\tonde{\frac{4\,\omega_{11}\,e^{2\omega_{11}t}+81\,\lambda_2^3\,\omega_{21}\,e^{2\omega_{21}t}}{4\,e^{2\omega_{11}t}+81\,\lambda_3^2e^{2\omega_{21}t}}}.\label{eq:ScalarEnergyFlux1121} \\
	\dot{J}_S&=&\frac{1}{\mu}\dot{E}_S.
\end{eqnarray}
The last relation is expected because both modes have $m=1$. 
In this case it is sufficient to average Eq.~\eqref{eq:ScalarEnergyFlux1121},
\begin{align*}
	\media{\dot{E}_S}&=\frac{M_S\,\mu}{2\pi}\quadre{\ln\tonde{4\,e^{\frac{4\pi\omega_{11}}{\mu}}+81\,\lambda_3^2\,e^{\frac{4\pi\omega_{21}}{\mu}}}-\ln\tonde{4+81\,\lambda_3^2}} \\
	\media{\dot{J}_S}&=\frac{1}{\mu}\media{\dot{E}_S}.
\end{align*}
Again, by expanding the above relations in the $M\mu\ll1$ limit, we finally obtain Eqs.~\eqref{eq:EdotS1121omegaIpiccolo}--\eqref{eq:JdotS1121omegaIpiccolo}.
%
\bibliographystyle{utphys}
\bibliography{Ref}
\end{document}